\documentclass[letter,11pt]{article}
\pdfoutput=1
\usepackage[utf8]{inputenc}

\makeatletter
\g@addto@macro\@floatboxreset\centering
\makeatother

\usepackage[paper=letterpaper,margin=1.3in]{geometry}

\usepackage{amsmath,amssymb,amsfonts,epsfig,cite,setspace,bigstrut,longtable,array,url,xcolor,microtype,mathtools,color}
\usepackage{bbm}                          % bbold 1
\definecolor{darkblue}{rgb}{0.0,0.0,0.4} 		% Dark-blue links 
\mathtoolsset{centercolon}
\usepackage{caption}
\usepackage{tikz}
\usetikzlibrary{backgrounds}
\usepackage{cancel}
\usepackage{subcaption}
\usepackage[unicode=true,
bookmarks=true,bookmarksnumbered=false,bookmarksopen=false,
breaklinks=false,pdfborder={0 0 0},pdfborderstyle={},backref=false,colorlinks=true]
{hyperref}
\hypersetup{pdftitle={From beta to eta: a new cohomology for deformed Sasaki--Einstein},
	pdfauthor={},
	linkcolor=darkblue, urlcolor=darkblue, citecolor=darkblue, filecolor=black, linktoc=page}

\usepackage[small]{titlesec}
\usepackage[framemethod=tikz]{mdframed}
\AtBeginEnvironment{mdframed}{%
	\tikzset{every picture/.style={}}%
}
\mdfsetup{roundcorner=.5ex}
{\begin{mdframed}[backgroundcolor=black!5!white]}%
	{\end{mdframed}}

\numberwithin{equation}{section}

\PassOptionsToPackage{linecolor=blue,backgroundcolor=blue!25,bordercolor=blue,textsize=scriptsize}{todonotes}
\usepackage{todonotes}

\allowdisplaybreaks

\usepackage{lipsum}

\usepackage{booktabs}       % nice tables 

\let\OLDthebibliography\thebibliography
\renewcommand\thebibliography[1]{
  \OLDthebibliography{#1}
  \setlength{\parskip}{0pt}
  \setlength{\itemsep}{2pt plus 0.3ex}
}

\makeatletter
\g@addto@macro\bfseries{\boldmath}
\makeatother

\let\originalleft\left
\let\originalright\right
\renewcommand{\left}{\mathopen{}\mathclose\bgroup\originalleft}
\renewcommand{\right}{\aftergroup\egroup\originalright}

\definecolor{purple}{cmyk}{0,0.8,0,0.4}
\definecolor{db}{cmyk}{1,0.1,0.2,0.6}
\definecolor{dg}{cmyk}{1,0,1,0.7}
\definecolor{bl}{cmyk}{1,0.7,0.5,0.2}
\definecolor{yl}{rgb}{0.2,0.7,0.2}
\definecolor{bl2}{cmyk}{0.7,0.4,0,0.5}
\definecolor{red2}{cmyk}{0,1,1,0.8}
\definecolor{red3}{cmyk}{0,0.7,1,0.7}
\definecolor{gr2}{cmyk}{1,0.2,0.7,0.6}
\definecolor{nb}{rgb}{0.1,0.1,0.5}
\definecolor{ng}{rgb}{0,0.8,0}
\definecolor{brown}{rgb}{0.6,0.3,0.2}
\definecolor{newred}{cmyk}{0,1,1,1}

\DeclareMathOperator{\vol}{vol}					% Volume
\DeclareMathOperator{\tr}{tr}					% Trace
\DeclareMathOperator{\image}{im}				% Image
\newcommand{\AdS}[1]{\text{AdS}_{#1}}			% AdS in math

\newcommand{\ext}{\mbox{\large $\wedge$}} 					% Exterior algebra
\newcommand{\dd}{\mathrm{d}} 								% Exterior derivative
\newcommand{\del}{\partial} 								% Exterior derivative
\newcommand{\delb}{\bar{\partial}} 								% Exterior derivative
\newcommand{\ii}{\mathrm{i}} 								% Imaginary number
\newcommand{\proj}[1]{\times_{#1}}							% Projection operator
\newcommand{\rep}[1]{\boldsymbol{#1}} 						% Group representations 
\newcommand{\id}{\boldsymbol{1}} 							% Identity element
\newcommand{\bbZ}{\mathbb{Z}} 								% Integers
\newcommand{\bbR}{\mathbb{R}} 								% Reals
\newcommand{\bbC}{\mathbb{C}} 								% Complex 
\newcommand{\cN}{\mathcal{N}}

\newcommand{\HC}{\mathit{HC}}                               % cyclic homology 
\newcommand{\rHC}{\overline{\HC}}                           % reduced cyclic homology 
\newcommand{\Ist}{\mathcal{I}_{\text{s.t.}}}                % single trace index

\newcommand{\TVindex}[1]{\mathrm{ind}_{\bdel}(#1)}          % transverse Dolbeault index
\newcommand{\etaindex}[1]{\mathrm{ind}_{\dd_{\eta}}(#1)}    % eta complex index

\newcommand{\dP}[1]{\mathrm{dP}_{#1}}

\newcommand{\db}{b}                                         % two-form in eta complex

\newcommand{\SU}[1]{\mathrm{SU}(#1)}
\newcommand{\Uni}[1]{\mathrm{U}(#1)}

\newcommand{\Ex}[1]{\mathrm{E}_{#1}}

\usepackage{tikz-cd}

\global\long\def\rep#1{\boldsymbol{#1}}%
\global\long\def\dd{\text{d}}%
\global\long\def\ii{\text{i}}%
\global\long\def\Uni#1{\text{U}(#1)}%
\global\long\def\SU#1{\text{SU}(#1)}%
\global\long\def\Ex#1{\text{E}_{#1}}%
\global\long\def\vol{\operatorname{vol}}%
\global\long\def\tr{\operatorname{tr}}%
\global\long\def\ext{\mbox{\large{\ensuremath{\wedge}}}}%
\global\long\def\image{\operatorname{im}}%
\global\long\def\op#1{\operatorname{#1}}%
\global\long\def\bR{\mathbb{R}}%
\global\long\def\bC{\mathbb{C}}%
\global\long\def\id{\operatorname{id}}%
\global\long\def\del{\partial}%
\global\long\def\bdel{\bar{\partial}}%
\global\long\def\kdel{\partial_{b}}%
\global\long\def\kbdel{\bar{\partial}_{b}}%

\global\long\def\nbr#1{\left( #1 \right)}%

\global\long\def\sbr#1{\left[ #1 \right]}%

\global\long\def\cbr#1{\left\{  #1 \right\}  }%

\global\long\def\om{\omega}%
\global\long\def\CRbundle{T_{1,0}}%
\global\long\def\CRanti{T_{0,1}}%
\global\long\def\CRquotient{\hat{T}}%

\global\long\def\Om{\Omega}%

\global\long\def\sg{\sigma}%

\global\long\def\KRd{\del_{b}}%

\global\long\def\KRdb{\delb_{b}}%

\global\long\def\TransverseSpace{\Lambda_{T}}%

\global\long\def\TransverseSpaceK#1{\TransverseSpace(#1)}%
\global\long\def\TransverseSpacerK#1#2{\TransverseSpace^{#1}(#2)}%

\global\long\def\TVSpace#1#2{\Lambda^{(#1,#2)}}%

\global\long\def\TVSpaceK#1#2#3{\TVSpace{#1}{#2}(#3)}%

\global\long\def\KRSpace#1#2{\Lambda^{[#1,#2]}}%

\global\long\def\KRSpaceK#1#2#3{\KRSpace{#1}{#2}(#3)}%

\global\long\def\etaSpace#1{\Lambda_{\eta}^{#1}}
\global\long\def\etaSpaceK#1#2{\Lambda_{\eta}^{#1}(#2)}

\global\long\def\TVH#1#2#3{H_{\bdel}^{(#1,#2)}(#3)}%
\global\long\def\KRH#1#2#3{H_{\kbdel}^{[#1,#2]}(#3)}%
\global\long\def\KRHnok#1#2{H_{\kbdel}^{[#1,#2]}}%
\global\long\def\etaH#1#2{H_{\dd_{\eta}}^{#1}(#2)}%
\global\long\def\tilden{\Sigma}%
\global\long\def\etaHnok#1{H_{\dd_{\eta}}^{#1}}%

\global\long\def\TVHarm#1#2#3{\mathcal{H}_{\Delta_{\bdel}}^{(#1,#2)}(#3)}%

\global\long\def\LieD{\mathcal{L}}%

\global\long\def\EllipticDk{D_{k}}%

\global\long\def\Ric{\text{Ric}}%

\global\long\def\TVh#1#2#3{h^{(#1,#2)}_{#3}}%
\global\long\def\KRh#1#2#3{h^{[#1,#2]}_{#3}}%

\let\oldlrcorner\lrcorner
\renewcommand\lrcorner{\oldlrcorner\,}

\begin{document}

\begin{titlepage}
\begin{flushright}Imperial/TP/21/ET/2\end{flushright}
\vfill
\begin{center}
{\setstretch{1.3}\Large\bf From $\beta$ to $\eta$: a new cohomology for \\ deformed Sasaki--Einstein manifolds\par} 
\vskip 1cm 
Edward Lødøen Tasker
\vskip .8cm
{\setstretch{1.3}\textit{Department of Physics, Imperial College London,\\ 
Prince Consort Road, London, SW7 2AZ, UK}\par} 
\end{center}
\vfill
\begin{center} \textbf{Abstract} \end{center}
\begin{quote} We discuss in detail the different analogues of Dolbeault cohomology groups on Sasaki--Einstein manifolds and prove a new vanishing result for the transverse Dolbeault cohomology groups $\TVH p0k$ graded by their charge under the Reeb vector. We then introduce a new cohomology, $\eta$-cohomology, which is defined by a CR structure and a holomorphic function $f$ with non-vanishing $\eta\equiv\dd f$. It is the natural cohomology associated to a class of supersymmetric type IIB flux backgrounds that generalise the notion of a Sasaki--Einstein manifold. These geometries are dual to finite deformations of the 4d $\cN=1$ SCFTs described by conventional Sasaki--Einstein manifolds. As such, they are associated to Calabi--Yau algebras with a deformed superpotential. We show how to compute the $\eta$-cohomology in terms of the transverse Dolbeault cohomology of the undeformed Sasaki--Einstein space. The gauge-gravity correspondence implies a direct relation between the cyclic homologies of the Calabi--Yau algebra, or equivalently the counting of short multiplets in the deformed SCFT, and the $\eta$-cohomology groups. We verify that this relation is satisfied in the case of S$^5$, and use it to predict the reduced cyclic homology groups in the case of deformations of regular Sasaki--Einstein spaces. The corresponding Calabi--Yau algebras describe non-commutative deformations of $\mathbb{P}^2$, $\mathbb{P}^1\times\mathbb{P}^1$ and the del Pezzo surfaces.

\end{quote}
\vfill
% footnote with emails and no number - tell hyperref not to throw an error
{\begin{NoHyper}\let\thefootnote\relax\footnotetext{\!\!\!\!\!\!\!\!\!Correspondence should be addressed to \tt d.waldram@imperial.ac.uk}\end{NoHyper}}
\end{titlepage}

\microtypesetup{protrusion=false}
\tableofcontents
\microtypesetup{protrusion=true}

%%%%%%%%%%%%%%%%%%%%%%%%%%%%

\section{Introduction}

The study of Sasaki--Einstein spaces plays a key role in string theory as they give the geometry underlying one of the canonical examples of the AdS/CFT correspondence~\cite{Maldacena:1997re}. For each five-dimensional Sasaki--Einstein space $M$ there is an equivalence between type IIB string theory in a spacetime that is asymptotically $\AdS5\times M$ (where $\AdS5$ is five-dimensional anti-de Sitter space) and a particular four-dimensional $\cN=1$ superconformal field theory (SCFT)~\cite{Kehagias:1998gn,Klebanov:1998hh,Acharya:1998db,Morrison:1998cs}. 

Sasaki--Einstein spaces can be defined by the condition that the metric cone over $M$ is Calabi--Yau. It turns out that many of the key properties of the dual $\mathcal{N}=1$ SCFTs depend only on holomorphic data, that is, on the complex structure on the cone. In particular, one can consider classes of operators in the field theory that transform in ``short multiplets'' of the $\mathcal{N}=1$ superconformal symmetry and are dual to Kaluza--Klein modes on the Sasaki--Einstein space. As shown by Eager, Schmude and Tachikawa~\cite{Eager:2012hx}, these are counted by the dimensions of particular cohomology groups on $M$. The \emph{Kohn--Rossi cohomology} of $M$, introduced in~\cite{KR65}, depends only on the CR structure on $M$, that is, the involutive subbundle $\CRbundle\subset TM\otimes\bbC$ defined by the complex structure on the cone. All Sasaki--Einstein metrics admit a Killing vector $\xi$, known as the Reeb vector, that generates the dual of the R-symmetry of the $\mathcal{N}=1$ SCFT. This can be used to refine the Kohn--Rossi cohomology to the \emph{transverse Dolbeault cohomology} groups $\TVH pqk$ graded by their charge $k$ under the action of the Reeb vector; it is the dimensions of these groups that count the Kaluza--Klein short multiplets.\footnote{Note that in~\cite{Eager:2012hx}, the groups $\TVH pqk$ are referred to as the Kohn--Rossi cohomologies, whereas more strictly they are the transverse cohomology groups. As we will discuss below, there is a direct relation between the two.}  

Mathematically, as discussed in~\cite{Eager:2012hx}, the dual $\mathcal{N}=1$ SCFT defines a \emph{Calabi--Yau algebra} $A$, first introduced by Ginzburg~\cite{Ginzburg:2006fu}. The archetypal construction of $A$ is from a quiver $Q$, encoding the fields of the SCFT, together with a superpotential $\mathcal{W}$. The short multiplets are then counted by the \emph{reduced cyclic homology} of the algebra $\rHC_n(A,k)$, graded by their R-charge $k$~\cite{Berenstein:2000ux,Berenstein:2001jr,Berenstein:2002fi,Eager:2012hx}. Explicitly, using the notation of~\cite{Eager:2012hx}, the chiral scalar (such as $\tr \mathcal{O}_f$), semi-conserved scalar ($\tr\mathcal{O}_v$) and semi-conserved $(0,1/2)$-spinor (such as $\tr \bar{W}_{\dot{\alpha}}\mathcal{O}_f$) multiplets are counted by the dimension of $\rHC_n(A,k)$, with $n=0,1,2$ respectively. The corresponding index 
\begin{equation}
    \Ist(t) = \sum_{0\leq n \leq 2,\ k>0} (-1)^n t^{2k}\dim\rHC_n(A,k)
\end{equation}
is known as the \emph{single-trace superconformal index} of the SCFT~\cite{Romelsberger:2005eg,Kinney:2005ej}. This index is independent of exactly marginal deformations of the field theory and can be extracted directly from the quiver description of the theory~\cite{Gadde:2010en}. 

In the special case where the SCFT is dual to $\AdS5$ times a Sasaki--Einstein manifold $M$, the Calabi--Yau algebra $A$ has the same cyclic homology as the coordinate ring of the cone over $M$. The reduced cyclic homology groups $\rHC_n(A,k)$ are then directly related to the transverse Dolbeault cohomology groups $\TVH pqk$, namely for $k>0$
\begin{equation}
\label{eq:rHC-TVH}
    \rHC_n(A,k) \simeq \sum_{p-q=n} \TVH pqk , 
\end{equation}
demonstrating the duality between counting operators in the field theory and Kaluza--Klein modes in the geometry~\cite{Eager:2012hx}. The index then takes the form 
\begin{equation}
    \Ist(t) = \sum_{k>0} \TVindex k t^{2k},
\end{equation}
where, by using vanishing properties of the $\TVH pqk$ groups, we can write
\begin{equation}
    \TVindex k \equiv \sum_{p,q} (-1)^{p+q} \dim \TVH pqk,
\end{equation}
which is the analogue of the Euler index at fixed $k$ for the transverse Dolbeault cohomology. 

There is a much larger class of SCFTs where the dual geometry is more complicated, involving many more of the fields in the type IIB supergravity than simply the metric and five-form that appear in the Sasaki--Einstein solution. Of particular interest are the theories that are exactly marginal deformations of those with Sasaki--Einstein duals, where the quiver $Q$ is unchanged but the superpotential $\mathcal{W}$ is modified. The canonical example is the set of $\mathcal{N}=1$ deformations of $\mathcal{N}=4$ super-Yang--Mills theory~\cite{Leigh:1995ep}, where the superpotential takes the form
\begin{equation}
\label{eq:N=4W}
\begin{split}
    \mathcal{W} &= h \tr\bigl(\Phi^1\Phi^2\Phi^3-\Phi^3\Phi^2\Phi^1\bigr)
        \\ & \qquad \qquad 
        + f_\beta \tr\bigl(\Phi^1\Phi^2\Phi^3+\Phi^3\Phi^2\Phi^1\bigr) 
        + f_\lambda \tr\bigl( (\Phi^1)^3 + (\Phi^2)^3 + (\Phi^3)^3 \bigr) .
\end{split}
\end{equation}
Setting $f_\beta=f_\lambda=0$ gives the $\mathcal{N}=4$ theory, where $A$ is simply the polynomial ring on $\bbC^3$ and the dual geometry is the five-sphere $M=\text{S}^5$. More generally $A$ is a non-commutative Sklyanin algebra (see for example~\cite{VdB92}). For $f_\lambda=0$, the dual type IIB background was derived in~\cite{Lunin:2005jy}. For general values of $f_\beta$ and $f_\lambda$, although the solutions lie in the class of backgrounds characterised in~\cite{Gauntlett:2005ww}, finding the explicit dual geometry has remained an open problem. Furthermore, one would expect there to be some new notion of cohomology, generalising the $\TVH pqk$ groups, that counts the number of short multiplets defined by the deformed non-commutative algebra.

The author and his collaborators have very recently given a solution to the first problem~\cite{paper}, finding the form of the supergravity background corresponding to an arbitrary finite exactly marginal deformation of \emph{any} field theory that is dual to a Sasaki--Einstein manifold. The analysis uses the formulation of the solution in terms of generalised geometry~\cite{Ashmore:2016qvs}. Somewhat in analogy to the case of Calabi--Yau manifolds, one first finds an explicit solution to a slightly weaker set of conditions (known as an “exceptional Sasaki” space) and then argues for the existence of the exact dual geometry using continuity. Crucially, there is a notion of holomorphic structure that is common to both the exceptional Sasaki space and the exact solution. In the dual field theory, this holomorphic structure encodes the superpotential, with the transition from the exceptional Sasaki space to the exact solution then viewed as a flow to the conformal fixed point. More precisely, the holomorphic structure is given by the CR structure of the Sasaki--Einstein geometry together with a function $f$ that is holomorphic on the Calabi--Yau cone and has charge three under the action of the Reeb vector. The function $f$ is the superpotential deformation $\Delta\mathcal{W}$ written as an element of the coordinate ring defined by the undeformed theory. For example, for the $\mathcal{N}=1$ deformations of $\mathcal{N}=4$ in~\eqref{eq:N=4W} one has 
\begin{equation}
    f=2f_\beta \, xyz + f_\lambda(x^3+y^3+z^3),
\end{equation}
where $(x,y,z)$ are complex coordinates on the cone $C(\text{S}^5)=\mathbb{C}^3$.

This paper is in part the companion to the work in the letter~\cite{letter} and has two main goals. The first is a review of Kohn--Rossi and transverse Dolbeault cohomologies in the context of Sasaki--Einstein manifolds, including some new results, such as a new bound on $\TVH p0k$. The second goal is to define new “$\eta$-cohomology” groups $\etaH nk$, where $\eta\equiv\dd f$ is assumed to be nowhere vanishing. These are a generalisation of the transverse Dolbeault cohomologies to the new “exceptional Sasaki--Einstein” geometries discussed in~\cite{paper}. They depend only on the holomorphic structure of the background and count short multiplets, hence they correspond to the reduced cyclic homology groups $\rHC_n(A,k)$ for the deformed non-commutative Calabi--Yau algebras $A$. Specifically we show that the AdS/CFT correspondence implies that~\eqref{eq:rHC-TVH} is replaced by 
\begin{equation}\label{eq:intro_relation}
    \rHC_n(A,k) \simeq \etaH {3-n}k , 
\end{equation}
for $k>0$. Furthermore, we show how to calculate the dimensions of $\etaH nk$ in terms of the $\TVH pqk$ groups of the undeformed theory. In particular, we show that $\etaH nk \simeq \etaH {4-n}{3-k}$ and, in all cases,\footnote{We use ``Iverson bracket'' notation $[S]$ that evaluates to 1 if the contained statement $S$ is true, and 0 if $S$ is false. In addition, $\equiv_3$ denotes equality modulo 3.}
\begin{equation}
    \begin{aligned}
        \dim \etaH 0k &= [k\equiv_3 0] , \\
        \dim \etaH 1k &= 0 , \\
        \dim \etaH 2k &= \TVindex k - [k\equiv_3 0] , 
    \end{aligned}
\end{equation}
thus giving a general prediction for the dimensions of $\rHC_n(A,k)$. In particular, we verify that \eqref{eq:intro_relation} is satisfied in the case of S$^5$, and use it to predict the reduced cyclic homology groups in the case of deformations of regular Sasaki--Einstein spaces. The corresponding Calabi--Yau algebras describe non-commutative deformations of $\mathbb{P}^2$, $\mathbb{P}^1\times\mathbb{P}^1$ and the del Pezzo surfaces.

The paper is organised as follows. We begin in Section \ref{KR_sec} with a review of two cohomologies that can be defined on any Sasaki manifold, namely Kohn--Rossi and transverse Dolbeault cohomologies. We then specialise to the case of Sasaki--Einstein manifolds and derive a new vanishing result for transverse Dolbeault cohomologies graded by charge under the Reeb vector. In Section \ref{sec:eta-complex} we define a new set of cohomology groups, $\eta$-cohomologies, that arise naturally in the context of deformations of Sasaki--Einstein solutions of type IIB string theory, and compute them in terms of the transverse Dolbeault cohomology of the undeformed Sasaki--Einstein manifold. In Section \ref{sec:indices}, we review how certain cyclic homology groups of Calabi--Yau algebras that appear in $\cN=1$ SCFTs are related to counting Kaluza--Klein modes in the dual $\AdS5$ supergravity background. We then describe how the $\eta$-cohomologies count these modes in the deformed Sasaki--Einstein solutions and use this to compute the corresponding cyclic homologies. We finish in Section \ref{sec:examples} with some examples where one can explicitly compute the $\eta$-cohomologies and compare to known field theory results.

\subsection*{Note added}
Edward Lødøen Tasker passed away in January 2020. He obtained the results in this work and wrote a draft of this paper during his PhD studies at Imperial College London. The paper has been edited for publication by A.~Ashmore and D.~Waldram.

\medskip

\noindent
Ed was a much-loved colleague and friend, and a gifted physicist and mathematician with a seemingly endless supply of puns and a knack for solving problems in unexpected ways. We miss him greatly. We hope sharing his work with others will add to his memory. (AA and DW.)

\section{Kohn--Rossi and transverse cohomologies}\label{KR_sec}

In this section we take $M$ to be a compact $(2n+1)$-dimensional manifold with Sasaki structure $(g,I,\sg,\xi)$~\cite{sasaki1960}.\footnote{For a review of Sasaki structures, we refer the reader to~\cite{Sparks:2007us,Sparks:2010sn,BG08}.} Here $\sg$ is the contact one-form, $\xi$ denotes the Reeb vector, $g$ is the Riemannian metric, and $I$ is the endomorphism that serves as an almost complex structure transverse to the orbits of $\xi$. In our conventions these satisfy the algebraic identities 
\begin{gather}
\imath_\xi \sigma = 1,\qquad \imath_\xi \omega = 0,\qquad I^{2}=-\id+\xi\otimes\sg,\\
\om(IX,IY)=\om(X,Y),\qquad g(X,Y)=\om(X,IY)+\sg(X)\sg(Y),
\end{gather}
where $\om\equiv\tfrac{1}{2}\dd\sg$ is the transverse Kähler form. The $+\ii$ eigenbundle $\CRbundle\subset TM\otimes\bbC$ of $I$ acting on the complexified tangent space defines a CR structure~\cite{chern1974,tanaka1962,1976131}. By definition, this means $\CRbundle\cap\overline{\CRbundle}=\{0\}$ and $\CRbundle$ is involutive under the Lie bracket, that is $[W,Z]\in\Gamma(\CRbundle)$ for all $W,Z\in\Gamma(\CRbundle)$. In addition, the Reeb vector and the transverse almost complex structure satisfy the ``K-contact'' condition $\LieD_{\xi}I=0$, where $\LieD_\xi$ is the Lie derivative. We fix an orientation on $M$ by choosing $\vol=-\sg\wedge\om^{n}/n!$ and denote the usual inner product on complex $p$-forms by 
\begin{equation}
\langle\alpha,\beta\rangle\equiv\int\alpha\wedge\overline{\star\beta}=\int\bar{\beta}^{\sharp}\lrcorner\alpha\vol,\label{eq:inner_product}
\end{equation}
where a superscript $\sharp$ denotes raising the indices of a form using the metric $g$ and $\star$ is the Hodge star. 

In the language of CR structures,\footnote{For a review see~\cite{CRbook}.} $\CRbundle$ is of \emph{hypersurface type}, meaning, given a point $x\in M$, the spaces 
\begin{equation}
    U_x \equiv \cbr{ \gamma \in T^*_xM \;\middle|\; \imath_{X} \gamma = 0,~\forall\, X\in(\CRbundle\oplus\overline{\CRbundle})_x }
\end{equation}
define a (real) line bundle $U\to M$. If $M$ is orientable, $U$ is trivial and admits global nowhere-vanishing sections. If there is some such section $\sigma$, such that the corresponding Levi form 
\begin{equation}
    L_\sigma(Z,\overline{W}) \equiv -\tfrac12 \ii\, \dd\sigma(Z,\overline{W}) , \quad W,Z\in \Gamma(\CRbundle)
\end{equation}
is positive definite, the CR structure is said to be \emph{strictly pseudo-convex}. In this case, $\sigma$ defines a unique vector $\xi$ satisfying $\imath_\xi\sigma=1$, $\imath_\xi \dd\sigma=0$. Finally, if there is a $\sigma$ such that the associated $\xi$ is holomorphic, that is $[\xi,Z]\in\Gamma(\CRbundle)$, $\forall Z\in\Gamma(\CRbundle)$, the pair $(\CRbundle,\xi)$ defines a \emph{normal strictly pseudo-convex CR structure}~\cite{Tanaka-book}, or equivalently a Sasaki structure~\cite[Corollary~2.10]{alekseevsky2015}. 

\subsection{Transverse Dolbeault cohomology}\label{sec:trans_cohomology}

We say that a complex $p$-form $\alpha$ is \emph{transverse} if $\imath_{\xi}\alpha=0$ and denote the space of transverse complex forms by $\TransverseSpace$. Using $I$ we can decompose further by \emph{type} to give 
\begin{align}
\TransverseSpace & \equiv\cbr{\alpha\in\Gamma(\Lambda^{\bullet}T^{*}M\otimes\bC)\;\middle|\;\imath_{\xi}\alpha=0},\\
\TVSpace pq & \equiv\cbr{\alpha\in\TransverseSpace\;\middle|\;I\cdot\alpha=-\ii(p-q)\alpha}. \label{eq:TVspace}
\end{align}
As these spaces are mutually orthogonal with respect to $\langle\cdot,\cdot\rangle$, we can restrict the inner product to these spaces. Since $\LieD_{\xi}$ is anti-Hermitian with respect to the above inner product, and by virtue of the K-contact condition, we can consider fixed-charge refinements of the above spaces which are in the kernel of $(\LieD_{\xi}-\ii k)$ for $k\in\bR$. We use the notation $\TransverseSpaceK k$ and $\TVSpaceK pqk$ respectively for these spaces:
\begin{align}
\TransverseSpaceK k & \equiv\cbr{\alpha\in\TransverseSpace\;\middle|\;\mathcal{L}_\xi \alpha = \ii k \alpha},\\
\TVSpaceK pqk & \equiv\cbr{\alpha\in\TVSpace pq\;\middle|\;\mathcal{L}_\xi \alpha = \ii k \alpha}.
\end{align}

Following \cite{Tievsky08}, for a transverse form $\alpha$ one can define a transverse exterior derivative $\dd_{T}$
\begin{equation}
\dd_{T}\alpha\equiv\dd\alpha-\sg\wedge\LieD_{\xi}\alpha,
\end{equation}
which decomposes as $\dd_{T}=\del+\bdel$ by virtue of the Sasaki conditions, where the \emph{transverse Dolbeault operators} are
\begin{equation}
\del\colon\TVSpace pq\to\TVSpace{p+1}q,\qquad\bdel\colon\TVSpace pq\to\TVSpace p{q+1}.
\end{equation}
Note that these operators are well defined only when acting on transverse forms. They satisfy the identities 
\begin{equation}
\del^{2}=0=\bdel^{2},\qquad\{\del,\bdel\}=-2\om\wedge\LieD_{\xi},
\end{equation}
and they distribute over wedge products as the usual exterior derivative does. In contrast with the Kohn--Rossi operators, which will be introduced in Section \ref{app:KR_cohomology}, these operators are genuinely complex conjugates of one another, $(\bdel\alpha)^*=\del\overline{\alpha}$.

The main objects of interest in the following subsection will be the \emph{transverse Dolbeault cohomology} groups $\TVH pqk$ (with $\bC$ coefficients) of the complex 
\begin{equation}
\cdots\xrightarrow{\;\bdel\;}\TVSpaceK p{q-1}k\xrightarrow{\;\bdel\;}\TVSpaceK pqk\xrightarrow{\;\bdel\;}\TVSpaceK p{q+1}k\xrightarrow{\;\bdel\;}\cdots
\end{equation}
As we will show below, these are finite dimensional, admit Hodge decompositions, and obey Serre dualities; similar statements will also hold for the \emph{Kohn--Rossi cohomologies} of Section~\ref{app:KR_cohomology}. In addition to the work of~\cite{Tievsky08}, these cohomology groups were studied in the (equivalent) context of normal strictly pseudo-convex CR structures by Tanaka in~\cite[Section 3]{Tanaka-book}. For $k=0$ they correspond to the more familiar basic Dolbeault cohomology groups (reviewed for example in~\cite{Sparks:2010sn}). 

\subsection{Transverse Laplacians, Hodge theory, and Serre duality}\label{sec:trans_laplace}

We now use Hodge theory to analyse the groups $\TVH pqk$, reproducing results of~\cite{Tanaka-book}. Using the inner product~\eqref{eq:inner_product} to define the adjoint and the Lefschetz operator $L\equiv\om\wedge{}$, one can show that 
\begin{equation}
\dd^{\dagger}L\beta-L\dd^{\dagger}\beta=\dd(I\cdot\beta)-I\cdot(\dd\beta)+2(n-r)\sg\wedge\beta
\end{equation}
for an arbitrary $r$-form $\beta$ (where $I\cdot$ is the standard endomorphism action on forms), which implies the transverse Kähler identities of \cite{Tievsky08} (see also~\cite{Stromenger10,Schmude:2013dua}). Using these identities, the three transverse Laplacians
\begin{equation}
\Delta_{T}\equiv\dd_{T}\dd_{T}^{\dagger}+\dd_{T}^{\dagger}\dd_{T},\qquad\Delta_{\del}\equiv\del\del^{\dagger}+\del^{\dagger}\del,\qquad\Delta_{\bdel}\equiv\bdel\bdel^{\dagger}+\bdel^{\dagger}\bdel,
\end{equation}
can be related via\footnote{Our conventions match those of \cite{Stromenger10}, differing from \cite{Tievsky08} by a factor of 2 in the last term.} 
\begin{equation}
\Delta_{T}\alpha=\Delta_{\bdel}\alpha+\Delta_{\del}\alpha=2\Delta_{\bdel}\alpha-2\ii(n-r)\LieD_{\xi}\alpha\label{eq:transverse_laplacian}
\end{equation}
for a transverse $r$-form $\alpha$. The action of the de Rham Laplacian $\Delta$ on a transverse form can be expressed as 
\begin{equation}
\Delta\alpha=\Delta_{T}\alpha-\LieD_{\xi}^{2}\alpha+4LL^{\dagger}\alpha+2\sg\wedge(L^{\dagger}\dd_{T}\alpha-\dd_{T}L^{\dagger}\alpha).
\end{equation}

Since $\Delta_{\bdel}$ commutes with both $I$ and $\mathcal{L}_{\xi}$, we can define the spaces of $\Delta_{\bdel}$-harmonics with $(p,q)$ type and fixed Reeb charge: 
\begin{equation}
\TVHarm pqk\equiv\cbr{\alpha\in\TVSpaceK pqk\;\middle|\;\Delta_{\bdel}\alpha=0}.
\end{equation}
As noted in \cite{Tievsky08}, whilst $\Delta_{\bdel}$ is \emph{not} elliptic on $\TransverseSpace$, the operator $2\Delta_{\bdel}-\LieD_{\xi}^{2}$ \emph{is} elliptic. The Hermitian operator 
\begin{equation}
\label{eq:Dk-def}
\EllipticDk\equiv\Delta_{\bdel}-\tfrac{1}{2}(\LieD_{\xi}-\ii k)^{2}
\end{equation}
will thus be elliptic on $\TransverseSpace$ for all $k\in\bR$, as its principal symbol is fixed and invertible. Consequently, we have the orthogonal decomposition 
\begin{equation}
\TransverseSpaceK k=\ker\EllipticDk\oplus\image\EllipticDk,
\end{equation}
and $\ker D_{k}$ is finite dimensional. Following the argument presented in \cite{Tievsky08}, which considers the $k=0$ case, if $\alpha\in\ker\EllipticDk$ it follows that $\Delta_{\bdel}\alpha=0$ and $(\LieD_{\xi}-\ii k)\alpha=0$, and thus, since $\Delta_{\delb}$ preserves type and charge, 
\begin{equation}
\TVHarm pqk=\ker\EllipticDk\big|_{\TVSpace pq}.
\end{equation}
Since $\ker D_{k}$ is finite dimensional, the spaces of $\Delta_{\bdel}$-harmonics are also finite dimensional. As noted by Tanaka~\cite{Tanaka-book}, the eigenvalues $k$ also form a discrete subset (without accumulation) of $\bbR$. 

Again, straightforwardly applying the argument of \cite{Tievsky08} to our case, it follows that one has the orthogonal decomposition 
\begin{equation}
\TransverseSpaceK k=\ker\Delta_{\bdel}\big|_{\TransverseSpaceK k}\oplus\image\Delta_{\bdel}\big|_{\TransverseSpaceK k}.
\end{equation}
Therefore, by the standard argument 
\begin{equation}
\TVH pqk\simeq\TVHarm pqk,
\end{equation}
so that every $\bdel$-closed, charge-$k$, type-$(p,q)$ class admits a unique $\Delta_{\bdel}$-harmonic charge-$k$ representative of the same type. 

In~\cite{Tievsky08}, a transverse Hodge operator is defined via $\star_{T}\alpha=\imath_{\xi}\star\alpha$, serving as an isomorphism between $(p,q)$-forms and $(n-q,n-p)$-forms. In terms of this, the adjoints of $\del$ and $\bdel$ can be expressed as 
\begin{equation}
\del^{\dagger}=\star_{T}\bdel\star_{T},\qquad\bdel^{\dagger}=\star_{T}\del\star_{T}.
\end{equation}
Using these operators and the observation that for a charge-$k$, $(p,q)$-form $\alpha$ one has $\star_{T}^{2}\alpha=(-1)^{p+q}\alpha$, it follows that 
\begin{equation}
\Delta_{\bdel}\alpha=0\quad\iff\quad\Delta_{\bdel}\overline{\star_{T}\alpha}=0.
\end{equation}
This implies a ``Serre duality'' for the transverse Dolbeault cohomology: 
\begin{equation}
\TVH pqk\simeq\TVH{n-p}{n-q}{-k}.\label{eq:serre_transverse}
\end{equation}
We can also prove a simple vanishing result. Taking $\alpha\in\TVHarm pqk$ and using (\ref{eq:transverse_laplacian}), one has 
\begin{equation}
2k(n-p-q)\langle\alpha,\alpha\rangle=\langle\alpha,\Delta_{T}\alpha\rangle=\langle\dd_{T}\alpha,\dd_{T}\alpha\rangle+\langle\dd_{T}^{\dagger}\alpha,\dd_{T}^{\dagger}\alpha\rangle\geq0,
\end{equation}
which implies the two conditions~\cite{Tanaka-book}
\begin{align}
\text{\ensuremath{k<0} and \ensuremath{p+q<n}} & \implies\TVH pqk=0, \label{eq:vanishk<0} \\
\text{\ensuremath{k>0} and \ensuremath{p+q>n}} & \implies\TVH pqk=0, \label{eq:vanishk>0}
\end{align}
related by the Serre duality \eqref{eq:serre_transverse} we gave above.

The \emph{basic} (chargeless, transverse Dolbeault) cohomology groups are given by setting $k=0$. We can easily recover two results: first, $\om$ is $\Delta_{\bdel}$-harmonic so generates a non-trivial class of $\TVH 110$; second, all chargeless $\Delta_{\bdel}$-harmonic functions are also necessarily $\Delta$-harmonic, and so $\TVH 000$ counts the number of connected components of $M$. If we define $\dd_B$ to be the restriction of $\dd_T$ to basic forms $\TransverseSpaceK 0$, then $\dd_B^2=0$ and one can define the basic cohomology groups $H_{\dd_B}^r(M)$. As discussed in~\cite{Tievsky08}, since $\Delta_T=2\Delta_{\bdel}$ on basic forms, one has a Hodge decomposition for basic cohomology groups
\begin{equation}
\label{eq:Hodge-decomp}
    H_{\dd_B}^r(M,\bbC) \simeq \bigoplus_{p+q=r} \TVH pq0 . 
\end{equation}
There is also a standard relation between basic and de Rham cohomologies~\cite{Tondeur-book} that follows from the short exact sequence of complexes induced by
\begin{equation}
    0 \xrightarrow{} \TransverseSpacerK r0 \xrightarrow{i} \Lambda^r T^*M \xrightarrow{\imath_\xi} \TransverseSpacerK {r-1}0 \xrightarrow{} 0  , 
\end{equation}
where $i$ is the inclusion map. By the ``zig-zag'' lemma, this gives a long exact sequence in cohomology 
\begin{equation}
\cdots\xrightarrow{[\wedge\omega]}H_{\dd_B}^r(M,\bC) \xrightarrow{[i]}H_{\dd}^r(M,\bC)  \xrightarrow{[\imath_{\xi}]}H_{\dd_B}^{r-1}(M,\bC)\xrightarrow{[\wedge\omega]}
H_{\dd_B}^{r+1}(M,\bC)\xrightarrow{[i]}\cdots
\end{equation}
where square brackets denote induced maps on cohomologies, and $[\wedge\om]$ arises from the chain map provided by wedging with $\om$. If $H^1_{\dd}(M,\bC)$ vanishes, as it does on any positive-scalar-curvature Einstein manifold, the long exact sequence implies both $H^1_{\dd_B}(M,\bC)\simeq 0$ and the short exact sequence 
\begin{equation}
0 \xrightarrow{} H_{\dd_B}^0(M,\bC)\xrightarrow{[\wedge\omega]}H_{\dd_B}^2(M,\bC) \xrightarrow{[i]}H_{\dd}^2(M,\bC)  \xrightarrow{} 0 ,
\end{equation}
implying 
\begin{equation}
\label{eq:deRham-basic2}
    H_{\dd}^2(M,\bbC) \simeq H_{\dd_B}^2(M,\bbC)/ \bbC[\omega] . 
\end{equation}

\subsection{Kohn--Rossi cohomology\label{app:KR_cohomology}}

Since Sasaki manifolds define a CR structure $\CRbundle\subset TM\otimes\bbC$, we can also consider the \emph{tangential Cauchy--Riemann}, or \emph{Kohn--Rossi}, operator $\kbdel$~\cite{KR65}. Writing $\CRanti=\overline{\CRbundle}$, involutivity of $\CRanti$ implies that one can define an operator $\kbdel\colon\Gamma(\Lambda^p\CRanti^*)\to\Gamma(\Lambda^{p+1}\CRanti^*)$ satisfying $\kbdel^2=0$. More generally, defining the quotient bundle $\CRquotient=(TM\otimes\bbC)/\CRanti$, one can consider sections of \emph{holomorphic} vector bundles~\cite{Tanaka-book,yau1981}
\begin{equation}
    \KRSpace pq \equiv\Gamma( \Lambda^p  \CRquotient^*\otimes \Lambda^q\CRanti^* ) 
\end{equation}
with $\kbdel \colon \KRSpace pq\to \KRSpace p{q+1}$ and $\kbdel^2=0$ so that it defines a complex. 

While the corresponding \emph{Kohn--Rossi cohomology} groups $\KRHnok pq$ can be defined for any CR structure~\cite{KR65}, if we have a strictly pseudo-convex CR structure (or more generally a non-degenerate CR structure) one can use harmonic theory to derive a Serre-type duality and bounds. These are the cohomologies discussed for example in \cite{Katmadas:2017gxn}, giving an $n+1$ by $n$ Hodge diamond. An important bound~\cite{KR65,Tanaka-book}, is that 
\begin{equation}
    \text{$\KRHnok pq$ is finite dimensional for any $q$ with $1\leq q\leq n-1$.}
\end{equation}

One can also use the Levi form to decompose the cotangent space as $TM\otimes\bbC=\bbC\xi\oplus\CRbundle\oplus\CRanti$ and hence identify $\CRquotient^*\simeq\bbC\sigma\oplus\CRbundle^*$. This in turn means we can identify $\KRSpace pq$ with the spaces of transverse forms
\[
\KRSpace pq\simeq\TVSpace pq\oplus\sg\TVSpace {p-1}q,
\]
where elements of $\sg\TVSpace {p-1}q$ are given by $\sigma\wedge\alpha$ with $\alpha\in \TVSpace {p-1}q$. We denote elements of this space using square brackets as ``$[p,q]$-forms''. Note that type $[0,q]$ is equivalent to type $(0,q)$. The Kohn--Rossi types provide a decomposition of the entire exterior algebra as follows. The exterior derivative can be decomposed by projecting appropriately, $\dd=\kbdel+\kdel$, where
\begin{equation}
\kdel\colon\KRSpace pq\to\KRSpace{p+1}q,\qquad\kbdel\colon\KRSpace pq\to\KRSpace p{q+1},
\end{equation}
are the \emph{Kohn--Rossi operators}. These behave like conventional Dolbeault operators in that
\begin{equation}
\kbdel^{2}=0=\KRd^{2},\qquad\{\kbdel,\kdel\}=0,
\end{equation}
but despite the notation they are \emph{not} complex conjugates of one another.\footnote{One can construct a fixed-charge lift that identifies the fixed-charge exterior differential algebra of $M$ with the exterior differential algebra of a certain bundle on the cone $\bR^{+}\times M$, the kernel of the antiholomorphic part of the homothetic vector field. The operators that give rise to $\kdel$ and $\kbdel$ are then the Dolbeault operators of the cone.} They can be characterised entirely in terms of the transverse Dolbeault operators $\del$ and $\delb$: if $\alpha$ is transverse, one has 
\begin{equation}
\KRd\alpha=\sg\wedge\LieD_{\xi}\alpha+\del\alpha,\qquad\kbdel\alpha=\bdel\alpha,\qquad\KRd\sg=0,\qquad\kbdel\sg=2\om.
\end{equation}

We will primarily be interested in the  Kohn--Rossi cohomology groups $\KRH pqk$ graded by $\xi$-charge. Denoting by $\KRSpaceK pqk$ the fixed charge type-$[p,q]$ space, one defines $\KRH pqk$ as the cohomologies of the complex
\begin{equation}
\dots\xrightarrow{\;\kbdel\;}\KRSpaceK p{q-1}k\xrightarrow{\;\kbdel\;}\KRSpaceK pqk\xrightarrow{\;\kbdel\;}\KRSpaceK p{q+1}k\xrightarrow{\;\kbdel\;}\dots
\end{equation}
To relate the Kohn--Rossi and transverse Dolbeault cohomologies, consider the commutative diagram
\begin{equation}             
\begin{tikzcd}[row sep = large]                       
{}                     & \vdots \arrow[d, "\bdel"']                     & \vdots \arrow[d, "\KRdb"']                     & \vdots \arrow[d, "\bdel"']                     & {} \\                       0                      \arrow[r]                     & \TVSpaceK{p}{q-1}{k}   \arrow[r, "i"] \arrow[d, "\bdel"']                     & \KRSpaceK{p}{q-1}{k}   \arrow[r, "\nbr{-1}^{q-1}\imath_\xi"] \arrow[d, "\KRdb"']                     & \TVSpaceK{p-1}{q-1}{k} \arrow[r] \arrow[d, "\bdel"']                     & 0 \\                                           0                      \arrow[r]                     & \TVSpaceK{p}{q}{k}     \arrow[r, "i"] \arrow[d, "\bdel"']                     & \KRSpaceK{p}{q}{k}     \arrow[r, "\nbr{-1}^{q}\imath_\xi"]  \arrow[d, "\KRdb"']                     & \TVSpaceK{p-1}{q}{k}   \arrow[r] \arrow[d, "\bdel"']                     & 0 \\                       {}                     & \vdots                     & \vdots                     & \vdots                     & {}             
\end{tikzcd}         
\end{equation}
where $i$ is the inclusion map, the rows are all short exact sequences, and the columns are all chain complexes. From the ``zig-zag'' lemma, it follows that there is a long exact sequence in cohomology, in particular for each $[p,q]$ there is an exact sequence
\begin{equation}
\TVH{p-1}{q-1}k\xrightarrow{[\wedge\omega]}\TVH pqk\xrightarrow{[i]}\KRH pqk\xrightarrow{[\imath_{\xi}]}\TVH{p-1}qk\xrightarrow{[\wedge\omega]}\TVH p{q+1}k,
\end{equation}
where square brackets denote induced maps on cohomologies, and $[\wedge\om]$ arises from the chain map provided by wedging with $\om$. This implies the short exact sequence
\begin{equation}
0\xrightarrow{\;}\op{coker}[\wedge\om]\big|_{\TVH{p-1}{q-1}k}\xrightarrow{\;}\KRH pqk\xrightarrow{\;}\ker[\wedge\om]\big|_{\TVH{p-1}qk}\xrightarrow{\;}0
\end{equation}
of $\bC$-modules, which must split; there must be an isomorphism 
\begin{equation}
\label{eq:KRH-TVH}
\KRH pqk\simeq\op{coker}[\wedge\om]\big|_{\TVH{p-1}{q-1}k}\oplus\ker[\wedge\om]\big|_{\TVH{p-1}qk}.
\end{equation}
Thus we confirm that the Kohn--Rossi cohomologies are also finite dimensional. 

The above expression simplifies for $k\neq0$: as observed in \cite{Eager:2013mua}, if a transverse charge-$k$ form $\alpha$ is $\bdel$-closed then $\alpha\wedge\om=\ii\bdel\del\alpha/(2k)$ is $\bdel$-exact, or equivalently $\image[\wedge\om]\simeq0$. Thus, for $k\neq0$ we have 
\begin{equation}
k\neq0\implies\KRH pqk\simeq\TVH pqk\oplus\TVH{p-1}qk.
\end{equation}
With more knowledge about $\image[\wedge\om]$ for basic cohomologies ($k=0$) we can obtain similar expressions. For instance, in the case of a connected manifold, $\TVH 000$ is one dimensional (generated by $1$) and the image of $[\wedge\om]$ acting on it is also one dimensional, generated by $\om$ (which will be non-trivial). Thus, for connected $M$,
\begin{align}
\KRH 10k &\simeq \begin{cases} 
  \TVH10k \oplus \TVH 00k & k\neq 0 , \\
  \TVH100 & k=0, 
  \end{cases} \\
\KRH 11k &\simeq \begin{cases} 
  \TVH11k \oplus \TVH 01k & k\neq 0 , \\
  \TVH110/\bC[\omega] \oplus \TVH 010 & k=0. 
  \end{cases} 
\end{align}

From Serre duality of the transverse Dolbeault cohomologies \eqref{eq:serre_transverse}, it follows that there is also a Serre-type duality of the Kohn--Rossi cohomologies~\cite{1905.03474}:
\begin{equation}
\KRH pqk\simeq\KRH{n+1-p}{n-q}{-k}.
\end{equation}
This is straightforward to show for $k\neq0$. For $k=0$ it follows from Lefschetz decomposition, which in the chargeless case is compatible with $\Delta_{\bdel}$-harmonicity.

\subsection{Transverse Dolbeault cohomology of Sasaki--Einstein manifolds}
\label{sec:SEcohom}

From here on, we specialise to the case where $M$ is a \emph{Sasaki--Einstein} manifold, so that there exists a nowhere-vanishing $(n,0)$-form $\Om$ satisfying~\cite{Bar:1993gpi,Martelli:2006yb,Sparks:2010sn}
\begin{equation}
\dd\Om=\ii(n+1)\sg\wedge\Om.
\end{equation}
The chain map provided by wedging with $\Om$, 
\begin{equation}             
\begin{tikzcd}[row sep=large]
\TVSpaceK{0}{q-1}{k} \arrow[r, "\bdel"] \arrow[d, "\wedge\Om"']                             
& \TVSpaceK{0}{q}{k} \arrow[r, "\bdel"] \arrow[d, "\wedge\Om"']                             
& \TVSpaceK{0}{q+1}{k} \arrow[d, "\wedge\Om"'] \\                                          
\TVSpaceK{n}{q-1}{k+n+1} \arrow[r, "\bdel"]                             
& \TVSpaceK{n}{q}{k+n+1} \arrow[r, "\bdel"]                             
& \TVSpaceK{n}{q+1}{k+n+1}                 
\end{tikzcd}
\end{equation}
has an inverse, and thus the induced map on cohomologies, denoted by $\sbr{\wedge\Om}$, is an isomorphism: 
\begin{equation}
\label{eq:Omega-iso}
[\wedge\Om]\colon\TVH 0qk\xrightarrow{\;\sim\;}\TVH nq{k+n+1}.
\end{equation}
This relates a pair of opposite edges of the Hodge diamond for different charges. Combining this with Serre duality in (\ref{eq:serre_transverse}), it follows that 
\begin{equation}
\TVH 0qk\simeq\TVH 0{n-q}{-n-1-k}.
\end{equation}
Note however that for $0<q<n$ the right-hand side of this is trivial for $-n-1<k$, as is the left-hand side for $k<0$. Thus one obtains the vanishing result 
\begin{equation}\label{eq:vanishing_0q}
\text{\ensuremath{0<q<n}}\implies\TVH 0qk\simeq0.
\end{equation}

In Section \ref{sec:bounds}, we will put further bounds on $\TVH p0k$ when $M$ is Sasaki--Einstein. In particular we show $\TVH p00\simeq 0$ for $p>0$. From the Hodge decomposition on basic cohomology~\eqref{eq:Hodge-decomp} together with~\eqref{eq:deRham-basic2}, we hence find the general relation
\begin{equation}
\label{eq:deRham-110}
    H_{\dd}^2(M,\bbC) \simeq \TVH110/ \bbC[\omega] . 
\end{equation}
where $H_{\dd}^2(M,\bbC)$ is the usual second de Rham cohomology group. 

In what follows, we will primarily be interested in five-dimensional Sasaki--Einstein spaces. The non-zero cohomology groups in this case are 
\begin{equation}
\label{eq:TVsummary}
\begin{gathered}
    \TVH 00k \simeq \TVH 20{k+3} \simeq \TVH 22{-k} \simeq \TVH 02{-k-3} ,  \\
    \TVH 10k \simeq \TVH 12{-k} ,\\
    \TVH 11k \simeq \TVH 11{-k},  
\end{gathered}
\end{equation}
where $\TVH 00k$ and $\TVH10k$ vanish for $k<0$. As we will see in the next section, we can actually derive a stronger constraint that $\TVH10k$ vanishes for $k\leq 3/2$. Using \eqref{eq:KRH-TVH}, the corresponding non-zero Kohn--Rossi groups are given by
\begin{equation}
\label{eq:KR-SE5-1}
\begin{gathered}
    \KRH 00k \simeq \KRH 30{k+3} \simeq \KRH 32{-k} \simeq \KRH 02{-k-3} , \\
    \KRH 10k \simeq \KRH 22{-k} , \\
    \KRH 20k \simeq \KRH 12{-k} , \\
    \KRH 11k \simeq \KRH 21{-k} , \\
\end{gathered}
\end{equation}
where
\begin{equation}
\label{eq:KR-SE5-2}
\begin{aligned}
    \KRH 00k &\simeq  \TVH 00k , \\
    \KRH 10k &\simeq \begin{cases} 
       \TVH10k \oplus \TVH 00k & k\neq 0 , \\
        0 & k=0 ,
        \end{cases} \\
    \KRH 20k &\simeq \TVH 20k \oplus \TVH 10k , \\
    \KRH 11k &\simeq \begin{cases} 
       \TVH11k & k\neq 0 , \\
        \TVH110/\bC[\omega] \simeq H_\dd^2(M,\bbC) & k=0 .
        \end{cases} 
\end{aligned}
\end{equation}

\subsection{An index on the transverse Dolbeault cohomology}
\label{sec:index}

In comparing with the single-trace superconformal index of the dual field theory, as we will see, a particular combination of transverse Dolbeault cohomology groups appears, namely 
\begin{equation}
\label{eq:TVindex}
    \TVindex k = \sum_{p,q}(-1)^{p-q} \dim \TVH pqk . 
\end{equation}
If the transverse Dolbeault complex were elliptic then this would correspond to the index of the complex 
\begin{equation}
\ldots\xrightarrow{\;\bdel\;}\bigoplus_{p+q=1}\Lambda^{(p,q)}(k)\xrightarrow{\;\bdel\;}\bigoplus_{p+q=2}\Lambda^{(p,q)}(k)\xrightarrow{\;\bdel\;}\bigoplus_{p+q=3}\Lambda^{(p,q)}(k)\xrightarrow{\;\bdel\;}\ldots
\end{equation}
where the charge is fixed to $k$. Instead, we can just view it as defined by 
\begin{equation}
    \TVindex k = \ker(D_{k},\Lambda^{\text{even}}(k))-\ker(D_{k},\Lambda^{\text{odd}}(k)) , 
\end{equation}
where $\Lambda^{\text{even}}(k)=\bigoplus_{p+q=\text{even}}\Lambda^{(p,q)}(k)$ and $\Lambda^{\text{odd}}(k)=\bigoplus_{p+q=\text{odd}}\Lambda^{(p,q)}(k)$, and $D_{k}$ is the generalised Laplacian defined in \eqref{eq:Dk-def}. 

From Serre duality \eqref{eq:serre_transverse}, we note that 
\begin{equation}
    \TVindex {-k} = \TVindex{k},
\end{equation}
and on a five-dimensional Sasaki--Einstein space
\begin{equation}
    \TVindex k = \dim \TVH 00k + \dim \TVH 11k + \dim \TVH 20k - \dim \TVH 10k  ,
\end{equation}
when $k>0$. 

\subsection{New bounds on \texorpdfstring{$\TVH p0k$}{H(delbar)(p,0)(k)}}
\label{sec:bounds}

In the following, we will need a sharp bound on when $\TVH 10k$ can be non-trivial. In this section, we will derive a set of new bounds on the charge $k$ for which $\TVH p0k$ with $p>0$ is non-trivial. In particular, for $n=2$ and $p=1$ it will imply 
\begin{equation}
\text{\ensuremath{n=2} and \ensuremath{k\leq\tfrac{3}{2}}}\implies\TVH 10k\simeq0.
\end{equation}

We can find such a bound by extending a standard technique for Einstein manifolds of positive scalar curvature, where one uses a B\"ochner identity to obtain a lower bound on the eigenvalues of the de Rham Laplacian~\cite{bochner1946}. Much of this follows Perrone~\cite{Perrone82} and builds on the work of Gallot--Meyer~\cite{GallotMeyer75}. Taking $M$ to be a $d$-dimensional, compact Riemannian manifold without boundary, we define the Riemann curvature $R$ and the Ricci curvature $\op{Ric}$ as
\begin{equation}
\begin{split}
R(X,Y)Z & =[\nabla_{X},\nabla_{Y}]Z-\nabla_{[X,Y]}Z\\
&=\nabla_{X,Y}^{2}Z-\nabla_{Y,X}^{2}Z+\nabla_{T(X,Y)}Z, \\
(R(X,Y)Z)^{a} & =R^{a}{}_{bcd}Z^{b}X^{c}Y^{d}\equiv R(X,Y)^{a}{}_{b}Z^{b},\\
\op{Ric}_{ab} & =R^{c}{}_{acb},
\end{split}
\end{equation}
where $T$ is the torsion of the connection $\nabla$, and $X$, $Y$ and $Z$ are vector fields. Specialising to the case where $\nabla$ is the Levi-Civita connection (and so is torsion-free), for a $r$-form $\alpha$ one has
\begin{equation}
\begin{split}
\nabla_{[a}\nabla_{b]}\alpha_{c_{1}\ldots c_{r}} & =-\tfrac{1}{2}r\,R^{d}{}_{[c_{1}|ab}\alpha_{d|c_{2}\ldots c_{r}]}, \\
(\dd^{\dagger}\alpha)_{a_{2}\ldots a_{r}} & =-\nabla^{b}\alpha_{ba_{2}\ldots a_{r}},\\
(\dd\alpha)_{a_{0}\ldots a_{r}} & =(r+1)\nabla_{[a_{0}}\alpha_{a_{1}\ldots a_{r}]}=\nabla_{a_{0}}\alpha_{a_{1}\ldots a_{r}}-r\nabla_{[a_{1}}\alpha_{|a_{0}|a_{2}\ldots a_{r}]}. 
\end{split}
\end{equation}
With these definitions, it is simple to show
\begin{equation}
\begin{split}
\Delta\alpha & \equiv(\dd\dd^{\dagger}+\dd^{\dagger}\dd)\alpha=-\op{div}\alpha-Q\alpha, \\
(\op{div}\alpha)_{a_{1}\ldots a_{r}} & =\nabla^{b}\nabla_{b}\alpha_{a_{1}\ldots a_{r}},\\
(Q\alpha)_{a_{1}\ldots a_{r}} & =-r\op{Ric}_{b[a_{1}}\alpha^{b}{}_{a_{2}\ldots a_{r}]}+\tfrac{1}{2}r(r-1)R_{b_{1}b_{2}[a_{1}a_{2}}\alpha^{b_{1}b_{2}}{}_{a_{3}\ldots a_{r}]}. 
\end{split}
\end{equation}
Note that the operator $Q$ is real and self-adjoint with respect to the standard inner product on $r$-forms (\ref{eq:inner_product}). For a function $h$, one has
\begin{equation}
\int_{M}\vol\op{div}h=-\int_{M}\vol\dd^{\dagger}\dd h\propto\int_{M}\dd\star\dd h=0,
\end{equation}
where we have used that $M$ is compact and without boundary. For an Einstein manifold with $\text{Ric}=\kappa g$, this implies
\begin{align}
0 & =\int_{M}\vol\op{div}(\bar{\alpha}^{\sharp}\lrcorner\alpha)=2\langle\alpha,\op{div}\alpha\rangle+2\int_{M}\vol|\nabla\alpha|^{2}\nonumber \\
\Rightarrow\quad\langle\alpha,\Delta\alpha\rangle & =\int_{M}\vol|\nabla\alpha|^{2}+\kappa r\langle\alpha,\alpha\rangle+\tau(\alpha),\label{eq:nabla_sq}
\end{align}
where we have introduced
\begin{equation}
\begin{split}
|\nabla\alpha|^{2} & =\frac{1}{r!}\nabla^{c}\bar{\alpha}^{a_{1}\ldots a_{r}}\nabla_{c}\alpha_{a_{1}\ldots a_{r}},\\
\tau(\alpha) & \equiv-\frac{1}{2(r-2)!}\int_{M}\vol R_{b_{1}b_{2}a_{1}a_{2}}\alpha^{b_{1}b_{2}}{}_{a_{3}\ldots a_{r}}\bar{\alpha}^{a_{1}\ldots a_{r}}.
\end{split}
\end{equation}
Note that $\tau(\alpha)=0$ for $r<2$. Now we need a lemma\footnote{There is a proof of this lemma in English in \cite{tachibana1976}.} of Gallot--Meyer~\cite{GallotMeyer75} in the form
\begin{equation}
\int_{M}\vol|\nabla\alpha|^{2}\geq\frac{1}{r+1}\langle\dd\alpha,\dd\alpha\rangle+\frac{1}{d-r+1}\langle\dd^{\dagger}\alpha,\dd^{\dagger}\alpha\rangle.
\end{equation}
Specialising to $d\geq2r$, one can write this in terms of the de Rham Laplacian as
\begin{equation}
\int_{M}\vol|\nabla\alpha|^{2}\geq\frac{d-2r}{(r+1)(d-r+1)}\langle\dd\alpha,\dd\alpha\rangle+\frac{1}{d-r+1}\langle\alpha,\Delta\alpha\rangle.
\end{equation}
Using the expression for $\int_{M}\vol|\nabla\alpha|^{2}$ from \eqref{eq:nabla_sq}, we can rearrange this to give a bound on the first non-zero eigenvalue of $\Delta$
\begin{equation}
\langle\alpha,\Delta\alpha\rangle\geq\frac{d-r+1}{d-r}\bigl(\kappa r\,\langle\alpha,\alpha\rangle+\tau(\alpha)\bigr)+\frac{d-2r}{(r+1)(d-r)}\langle\dd\alpha,\dd\alpha\rangle.
\end{equation}

Taking $d=2n+1$ and $\kappa=2n$, for a general $r$-form $\alpha$ on a $(2n+1)$-dimensional Sasaki--Einstein manifold, where $\Ric=2ng$~\cite{Sparks:2010sn}, the above bound is
\begin{equation}
\langle\alpha,\Delta\alpha\rangle  \geq\frac{2n+2-r}{2n+1-r}\bigl(2nr\,\langle\alpha,\alpha\rangle+\tau(\alpha)\bigr)+\frac{2n+1-2r}{(r+1)(2n+1-r)}\langle\dd\alpha,\dd\alpha\rangle.
\end{equation}
Taking $r=1$, which implies $\tau(\alpha)=0$, and dropping the $\langle\dd\alpha,\dd\alpha\rangle$ term, we recover the standard lower bound for the Laplacian eigenvalue of a one-form, which implies in particular that the first de Rham cohomology is trivial for positive-scalar-curvature Einstein manifolds~\cite{bochner1946}.

Now suppose $\alpha$ to be a $\Delta_{\bdel}$-harmonic $(p,0)$-form of charge $k\geq0$ for $0<p\leq n$. Stromenger~\cite{Stromenger10} shows that the curvature of a Sasaki--Einstein metric satisfies
\begin{equation}
    R(U,V)Z = g(U,Z)V  - g(V,Z)U
\end{equation}
for all $U,V\in\Gamma(\CRbundle)$ and $Z\in\Gamma(TM)$. This implies for a $(p,0)$-form $\alpha$
\begin{equation}
    \tau(\alpha) = -p(p-1)\langle\alpha,\alpha\rangle .
\end{equation}
We also have 
\begin{align}
\langle\dd\alpha,\dd\alpha\rangle & =k^{2}\langle\sg\wedge\alpha,\sg\wedge\alpha\rangle+\langle\dd_{T}\alpha,\dd_{T}\alpha\rangle\geq k^{2}\langle\alpha,\alpha\rangle,\\
\langle\alpha,\Delta\alpha\rangle & =(k^{2}+4nk)\langle\alpha,\alpha\rangle.
\end{align}
If $\langle\alpha,\alpha\rangle\neq0$, it follows that 
\begin{equation}\label{eq:quad-ineq}
k^{2}+2k\frac{(n-p)(p+1)(2n-p+1)}{p(2n-p+2)}-(p+1)(2n-p+1)\geq0.
\end{equation}
This means that $\alpha$ can be non-trivial only for $k\geq k_+$, where 
\begin{equation}
    k_+ = \frac{(n-p)(p+1)(2n-p+1)}{p(2n-p+2)} \left[
          \sqrt{ \frac{p^2(2n-p+2)^2}{(n-p)^2(p+1)(2n-p+1)} + 1} - 1\right] 
\end{equation}
is the positive root in the quadratic inequality~\eqref{eq:quad-ineq}. Hence, given $\TVH p0k\simeq0$ for $k<0$, we have 
\begin{equation}
    k < k_+ \implies\TVH p0k\simeq0.
\end{equation}
Note that $k_+>0$ and so in particular we have $\TVH p00\simeq0$. For $p=n$ we have $k_+=n+1$. Given the isomorphism \eqref{eq:Omega-iso} and the facts that $\TVH00k\simeq0$ for $k<0$ and $\TVH000\simeq\bbC$, we see that in this case the bound is saturated.  

Of particular interest for us is the case $n=2$, so that $M$ is five dimensional. Taking $p=1$, we  have $k_+=(2\sqrt{66}-8)/5\approx1.6496$ and so, in conclusion, we have the triviality result 
\begin{equation}
\text{\ensuremath{n=2} and \ensuremath{k\leq\tfrac{3}{2}}}\implies\TVH 10k\simeq0.
\end{equation}

\section{The \texorpdfstring{$\eta$}{eta}-complex}
\label{sec:eta-complex}

As discussed in the introduction, there is a natural string theory extension of a five-dimensional Sasaki--Einstein manifold that describes a generic supersymmetric type IIB background of the form $\AdS5\times M$~\cite{Gauntlett:2005ww}. Using generalised geometry, one can identify a structure, known as the H-structure, that encodes the holomorphic information about the dual field theory~\cite{Ashmore:2016qvs}. In particular, as shown in the companion letter~\cite{letter} to this paper, for backgrounds that correspond in the field theory to marginal deformations of a SCFT dual to a Sasaki--Einstein geometry, the holomorphic structure is determined by the CR structure of the Sasaki--Einstein geometry and a holomorphic function $f$. In this section, we define a natural set of cohomology groups $\etaH sk$ defined by this generalised holomorphic structure and show how these are determined in terms of the transverse Dolbeault cohomology groups of the underlying Sasaki--Einstein manifold. In the following sections, we will show how they are related to the reduced cyclic homology groups $\rHC_n(A,k)$ of the Calabi--Yau algebras $A$ that describe the dual SCFTs and give some examples. 

The complex that defines the $\etaH sk$ cohomology is defined using the exact one-form
\begin{equation}
\eta \equiv \dd f = \kdel f,
\end{equation}
where $\kdel$ is a Kohn--Rossi differential, and where the holomorphicity condition on the function $f$ is 
\begin{equation}
\bdel f\equiv\kbdel f=0.
\end{equation}
We will make the additional assumption that $\eta$ is \emph{nowhere vanishing},\footnote{The condition that $\eta$ is nowhere vanishing is somewhat restrictive. Recall that Sasaki--Einstein spaces can be \emph{quasi-regular} or \emph{irregular}, depending on whether the orbits of the Reeb vector field $\xi$ are compact (and hence define a locally free $\Uni1$ action on $M$) or are non-compact. Our expectation is that the existence of a nowhere-vanishing $\eta$ implies that the undeformed Sasaki--Einstein is quasi-regular, though we have not been able to prove this.} which means that we also have a complex vector field $n$ that satisfies
\begin{equation}
\imath_{n}\eta=1.
\end{equation}
Given some real metric on the underlying manifold, such a vector field can always be constructed from $\eta$ as $n=(\bar{\eta}^{\sharp}\lrcorner\eta)^{-1}\bar{\eta}^{\sharp}$, where $\sharp$ indicates raising an index with the metric. 

Since it is non-vanishing, $\eta$ defines a subbundle of the tangent bundle $\mathcal{F}_{\eta}\hookrightarrow TM\otimes\bbC$ as
\begin{equation}
\mathcal{F}_{\eta}=\ker\eta=\{v\in\Gamma(TM\otimes\bbC)\;|\;\imath_{v}\eta=0\}.
\end{equation}
Since $\eta$ is closed by definition, the subbundle $\mathcal{F}_{\eta}$ is closed under the Lie bracket, that is $[v,w]\in\mathcal{F}_{\eta}$ for all $v,w\in\mathcal{F}_{\eta}$. As such it defines a complex Lie algebroid, and hence there is an associated differential $\dd_{\eta}$ acting on sections of  $\ext^{s}\mathcal{F}_{\eta}^{*}$, with the corresponding cohomology groups $H_{\dd_{\eta}}^{s}$. It is not difficult to check that sections of $\ext^{s}\mathcal{F}_{\eta}^{*}$ can be viewed as an equivalence class of complex $s$-forms: given an $s$-form $\alpha$ and $\beta$ any $(s-1)$-form, we identify
\begin{equation}
\label{eq:eta-equiv}
\alpha\sim\alpha+\eta\wedge\beta.
\end{equation}
Equivalently, we can identify the class of $\alpha$ with the $(s+1)$-form $\eta\wedge\alpha$, that is, we have an isomorphism 
\begin{equation}
\label{eq:eta-iso}
    \etaSpace s \equiv\Gamma(\eta\wedge(\ext^sT^*M\otimes\bbC)) 
    \simeq \Gamma(\ext^s \mathcal{F}_\eta^*) ,
\end{equation}
such that, given $\alpha$ in the equivalence class~\eqref{eq:eta-equiv}, 
\begin{equation}
\label{eq:eta-d}
    \dd_\eta \alpha \mapsto \dd(\eta\wedge\alpha) = \eta\wedge\dd\alpha .
\end{equation}
Thus, for example, $\alpha$ is $\dd_{\eta}$-closed if and only if $\eta\wedge\alpha$ is $\dd$-closed. 

With this in mind, consider the complex
\begin{equation}
\label{eq:eta_complex}
    \cdots\xrightarrow{\phantom{\;\dd\;}}
    \etaSpace{s-1}
    \xrightarrow{\;\dd\;}
    \etaSpace{s}
    \xrightarrow{\;\dd\;}
    \etaSpace{s+1}
    \xrightarrow{\phantom{\;\dd\;}}\cdots
\end{equation}
Given the identifications~\eqref{eq:eta-iso} and~\eqref{eq:eta-d}, one can compute the Lie algebroid cohomologies using the above complex. We will refer to these as ``$\eta$-cohomologies'', given by
\begin{equation}
    \etaHnok s=\frac{\ker\dd|_{\etaSpace{s}}}{\image\dd|_{\etaSpace{s-1}}}
.       =\frac{\{\dd\text{-closed }\eta\wedge\alpha_{(s)}\}}{\{\eta\wedge\dd\alpha_{(s-1)}\}}
    =\frac{\ker\dd_{\eta}|_{\mbox{\small $\wedge$}^{s}\mathcal{F}_{\eta}^{*}}}{\image\dd_{\eta}|_{\mbox{\small $\wedge$}^{s-1}\mathcal{F}_{\eta}^{*}}}.
\end{equation}

Recall that we are actually interested in the case where $\eta$ encodes a deformation of the holomorphic structure of a five-dimensional compact Sasaki--Einstein space.\footnote{Equivalently, a compact normal strictly pseudo-convex CR Einstein manifold~\cite{Tanaka-book}.} In this case $\eta$ has charge $+3$ under the action of the Reeb vector 
\begin{equation}
\mathcal{L}_{\xi}\eta=3\ii\eta.
\end{equation}
One can then grade the complex~\eqref{eq:eta_complex} by charge under the Reeb vector action. The differential $\dd_{\eta}$ (or $\dd$) commutes with $\mathcal{L}_{\xi}$ so we can restrict~\eqref{eq:eta_complex} to fixed charge $k$. In particular, we can define
\begin{equation}
    \etaSpaceK sk \equiv \cbr{\eta\wedge\alpha\in\etaSpace s\;\middle|\;\mathcal{L}_\xi (\eta\wedge\alpha) = \ii k (\eta\wedge\alpha)},
\end{equation}
implying $\alpha$ has charge $k-3$.  With this assignment, the charge-$k$ complex is given by
\begin{equation}
    \label{eq:eta_complex_k}
    \ldots\xrightarrow{\phantom{\;\dd\;}}
    \etaSpaceK{s-1}k
    \xrightarrow{\;\dd\;}
    \etaSpaceK{s}k
    \xrightarrow{\;\dd\;}
    \etaSpaceK{s+1}k
    \xrightarrow{\phantom{\;\dd\;}}\ldots
    % 0\xrightarrow{\phantom{\;\dd\;}}\Gamma(\eta\, T^{*})(k)\xrightarrow{\;\dd\;}\Gamma(\eta\,\ext^{2}T^{*})(k)\xrightarrow{\;\dd\;}\Gamma(\eta\,\ext^{3}T^{*})(k)\xrightarrow{\phantom{\;\dd\;}}\ldots
\end{equation}
with the corresponding graded  $\eta$-cohomology groups $\etaH sk$.

In the rest of this section, we will first show that there is a natural pairing that relates $\etaH sk\simeq \etaH {4-s}{-k}$ and then calculate $\etaH 2k$ in terms of the Kohn--Rossi (or equivalently transverse Dolbeault) cohomology groups of the underlying Sasaki--Einstein manifold. We then extend this result to $\etaH 0k$ and $\etaH 1k$. 

\subsection{Duality for \texorpdfstring{$\etaH sk$}{H(d eta)(s)(k)}}

We now want to introduce a pairing on the $\eta$-cohomology and prove a simple duality for the cohomology groups. Consider a pairing
\begin{equation}
\langle\eta\wedge\alpha,\eta\wedge\beta\rangle_{\eta}\equiv\int\eta\wedge\alpha\wedge\beta=\int(\eta\wedge\alpha)\wedge\imath_{n}(\eta\wedge\beta),
\end{equation}
where $\eta$ is again exact, nowhere vanishing and charge $+3$, and $\alpha$ and $\beta$ are two-forms. Taking $\eta\wedge\alpha$ and $\eta\wedge\beta$ to have fixed charges $k_{\alpha}$ and $k_{\beta}$ under the action of $\xi$, the pairing vanishes trivially if $k_{\alpha}+k_{\beta}\neq3$. Thus, we can take $k_{\alpha}=k$ and $k_{\beta}=3-k$ to focus on non-vanishing pairings.

Consider what happens when both $\alpha$ and $\beta$ are $\dd_{\eta}$-closed, so that $\eta\wedge\alpha$ and $\eta\wedge\beta$ are $\dd$-closed. It is then simple to show that the pairing does not depend on the representative of the $\eta$-cohomology classes. Taking $\alpha=\dd\gamma$, we have
\begin{equation}
\begin{split}
\langle\eta\wedge\dd\gamma,\eta\wedge\beta\rangle_{\eta}&=\int\eta\wedge\dd\gamma\wedge\beta
\\&=-\int\dd(\eta\wedge\gamma\wedge\beta)-\int\gamma\wedge\dd(\eta\wedge\beta)=0,
\end{split}
\end{equation}
where we have used compactness and Stokes' theorem. From this we see that the pairing is well defined on the classes. We can go further and prove that the pairing is actually non-degenerate on the cohomology. Non-degeneracy is the statement that if $\langle\eta\wedge\alpha,\eta\wedge\beta\rangle_{\eta}=0$ for all $\dd$-closed $\eta\wedge\beta$ of charge $3-k$, then there exists a charge-$(k-3)$ one-form $\gamma$ such that $\eta\wedge\alpha=\eta\wedge\dd\gamma$.

Let us take $\eta\wedge\alpha$ to be $\dd$-closed and of charge $k$ (so that $\alpha$ is charge $k-3$), and consider an ``action''
\begin{equation}
S[\eta\wedge\beta]=\langle\eta\wedge\alpha,\eta\wedge\beta\rangle_{\eta},
\end{equation}
where $\eta\wedge\beta$ is $\dd$-closed and charge $3-k$. Suppose that $\eta\wedge\beta_{*}$ extremises this action so that its first-order variation vanishes:
\begin{equation}
0=S[\eta\wedge(\beta_{*}+\delta\beta)]-S[\eta\wedge\beta_{*}]=\langle\eta\wedge\alpha,\eta\wedge\delta\beta\rangle_{\eta},
\end{equation}
where again $\eta\wedge\delta\beta$ is $\dd$-closed and charge $3-k$. This means that at the extrema of $S$, $\langle\eta\wedge\alpha,\eta\wedge\delta\beta\rangle_{\eta}$ vanishes for all $\delta\beta$. We would now like to prove that at these extrema, there must exist a one-form $\gamma$ with the properties mentioned above. Consider a related action where $\gamma$ is thought of as a Lagrange multiplier that imposes the constraint $\dd(\eta\wedge\beta)=0$:
\begin{equation}
S'[\eta\wedge\beta,\gamma]=S[\eta\wedge\beta]-\int\gamma\wedge\dd(\eta\wedge\beta),
\end{equation}
where $\gamma$ and $\beta$ are unconstrained other than having fixed charge. The extrema of $S'$ should match the extrema of $S$ under the constrained variations. Varying $S'$ around $\eta\wedge\beta_{*}$ and $\gamma_{*}$ to first order, we have
\begin{equation}
\begin{split}
0 & =S'[\eta\wedge(\beta_{*}+\delta\beta),\gamma_{*}+\delta\gamma]-S'[\eta\wedge\beta_{*},\gamma_{*}] \\
& =\int\eta\wedge\alpha\wedge\delta\beta-\int\gamma_{*}\wedge\dd(\eta\wedge\delta\beta)-\int\delta\gamma\wedge\dd(\eta\wedge\beta_{*})\\
& =\int\eta\wedge(\alpha-\dd\gamma_{*})\wedge\delta\beta-\int\delta\gamma\wedge\dd(\eta\wedge\beta_{*}). 
\end{split}
\end{equation}
For this to vanish for all $\delta\beta$ and $\delta\gamma$, we must have
\begin{equation}
\eta\wedge\alpha=\eta\wedge\dd\gamma_{*},\qquad\dd(\eta\wedge\beta_{*})=0.
\end{equation}
We see that at the extremum, $\eta\wedge\beta_{*}$ is $\dd$-closed and there exists a one-form $\gamma_{*}$ such that $\eta\wedge\alpha=\eta\wedge\dd\gamma_{*}$, implying the pairing is non-degenerate.

As the pairing is non-degenerate on the $\eta$-cohomology and pairs charge-$k$ with charge-$(3-k)$ elements, the corresponding cohomologies at charge-$k$ and charge-$(3-k)$ are isomorphic. % In terms of the numbers $n_{s,k}$ that count the $\eta$-cohomology, 
This is simply the statement that
\begin{equation}
% n_{s,k}=n_{4-s,3-k}.
    \etaH sk \simeq \etaH {4-s}{3-k} .
    \label{eq:charge_sym}
\end{equation}
This follows from repeating the previous calculation for charge-$k$ forms of different rank: for example, if $\alpha$ is a $s$-form, $\beta$ would be a $(4-s)$-form.

\subsection{Calculating \texorpdfstring{$\etaH 2k$}{H(d eta)(2)(k)}}

We now want to relate the charge-$k$ $\eta$-cohomologies $\etaH sk$ to the Kohn--Rossi (or equivalently transverse Dolbeault) cohomologies of the underlying Sasaki--Einstein manifold and the properties of $\eta=\dd f$. We start with $s=2$, as it is the most involved, and then turn to the other cases. 

Thanks to the observation in \eqref{eq:charge_sym}, we can restrict our attention to $k\geq3/2$. Recall that the relevant complex is \eqref{eq:eta_complex_k}. The charge-$k$, $s=2$ cohomology is then the cohomology of
\begin{equation}
    \etaSpaceK 1k \xrightarrow{\;\dd\;} \etaSpaceK 2k \xrightarrow{\;\dd\;}0,
    \label{eq:reduced_complex_0}
\end{equation}
that is we want to count the number of $\dd$-closed forms in $\etaSpaceK 2k$ modulo $\dd$-exact ones. We will denote elements of these spaces by
\begin{equation}
\eta\wedge\lambda\in\etaSpaceK 1k,\qquad\eta\wedge\db\in\etaSpaceK 2k.
\end{equation}
The key to computing the cohomology is to split the exterior derivative into the Kohn--Rossi operators $\dd=\partial_{b}+\bar{\partial}_{b}$, with a corresponding decomposition of forms into $[p,q]$ types. Under these conventions, $\eta=\dd f=\partial_{b}f$ is type $[1,0]$. The complex \eqref{eq:reduced_complex_0} then splits into
\begin{equation}\label{eq:double_complex}
\begin{tikzcd}[column sep={4em,between origins}, row sep={4em,between origins}]
& & \eta\wedge\lambda_{[0,1]} \arrow[rd, "\partial_b"] \arrow[ld, "\bar\partial_b"'] &  & \eta\wedge\lambda_{[1,0]} \arrow[rd, "\partial_b"] \arrow[ld, "\bar\partial_b"'] & & \\ 
& \eta\wedge\db_{[0,2]} \arrow[rd, "\partial_b"] \arrow[ld, "\bar\partial_b"'] &  & \eta\wedge\db_{[1,1]} \arrow[rd, "\partial_b"] \arrow[ld, "\bar\partial_b"'] & & \eta\wedge\db_{[2,0]} \arrow[rd, "\partial_b"] \arrow[ld, "\bar\partial_b"'] & \\
0 & & 0 & & 0 & & 0
\end{tikzcd}
\end{equation}
where we have denoted the $[p,q]$ type of each component with subscripts. Our plan is to proceed from left to right, imposing that the relevant forms are $\kdel$- or $\kbdel$-closed and then quotienting by exact forms.

We begin by noting that for any $\dd$-closed $\eta\wedge\db$, the component due to $\db_{[0,2]}$ is trivially $\bar{\partial}_{b}$-closed (or equivalently $\bar{\partial}$-closed as it is type $(0,2)$). By the lower bound on the charge of non-zero $\bar{\partial}_{b}$-closed functions and the various dualities we have already mentioned, $H_{\bar{\partial}_{b}}^{[0,2]}(k')$ is trivial for $k'>-3$. Since $\db_{[0,2]}$ has charge $k-3$ and we are restricting to $k\geq3/2$, $\db_{[0,2]}$ has charge greater than or equal to $-3/2$ and so can always be written as $\db_{[0,2]}=\bar{\partial}_{b}\mu_{[0,1]}$. This can always be shifted away using the freedom in $\lambda_{[0,1]}$ and so without loss of generality we can pick a representative with $\db_{[0,2]}=0$.

Note that we have not used up all of the freedom in $\lambda_{[0,1]}$ -- we can still shift by $\eta\wedge\dd\lambda_{[0,1]}$ provided $\eta\wedge\bar{\partial}\lambda_{[0,1]}=0$, or equivalently $\bar{\partial}\lambda_{[0,1]}=0$ (since $\imath_{n}\eta=1$ and $\imath_{n}$ annihilates $\Lambda^{[0,\bullet]}$). Given that $H_{\bar{\partial}}^{(0,1)}(k)$ is trivial on a compact connected five-dimensional Sasaki--Einstein manifold (see \eqref{eq:vanishing_0q}) and $\Lambda^{[0,\bullet]}=\Lambda^{(0,\bullet)}$, a $\delb$-closed $\lambda_{[0,1]}$ must be $\bar{\partial}_{b}$-exact and so can be written as $\lambda_{[0,1]}=\bar{\partial}_{b}\alpha_{[0,0]}$. Using $\{\partial_{b},\bar{\partial}_{b}\}=0$, we then have
\begin{equation}
\eta\wedge\dd\lambda_{[0,1]}=\eta\wedge\partial_{b}\bar{\partial}_{b}\alpha_{[0,0]}=\eta\wedge\dd(-\partial_{b}\alpha_{[0,0]}),
\end{equation}
implying that modding out by $\eta\wedge\dd\lambda_{[0,1]}$ with $\bar{\partial}\lambda_{[0,1]}=0$ is equivalent to modding out by some $\eta\wedge\dd\lambda_{[1,0]}$. Said differently, modding out by $\eta\wedge\dd\lambda_{[1,0]}$ alone is sufficient since this includes all possible $\eta\wedge\dd\lambda_{[0,1]}$ for $\bar{\partial}$-closed $\lambda_{[0,1]}$. This is already taken care of by the right-most part of the double complex \eqref{eq:double_complex}, so we can instead focus on
\begin{equation}\label{eq:reduced_complex}
\begin{tikzcd}[column sep={4em,between origins}, row sep={4em,between origins}]
&  & \eta\wedge\lambda_{[1,0]} \arrow[rd, "\partial_b"] \arrow[ld, "\bar\partial_b"'] & &  \\
& \eta\wedge\db_{[1,1]} \arrow[rd, "\partial_b"] \arrow[ld, "\bar\partial_b"'] & & \eta\wedge\db_{[2,0]} \arrow[rd, "\partial_b"] \arrow[ld, "\bar\partial_b"'] & \\
0 & & 0 & & 0
\end{tikzcd}
\end{equation}

To tackle this, we note first that since $\eta$ is nowhere vanishing and type $[1,0]$, we can write $\eta\wedge\db_{[2,0]}=\rho_{[3,0]}$. The strategy is then to parametrise the most general $\bar{\partial}_{b}$-closed $\eta\wedge\db_{[1,1]}$ and then mod out by $\eta\wedge\bar{\partial}_{b}\lambda_{[1,0]}$. The remaining freedom is shifts by $\kbdel$-closed $\eta\wedge\lambda_{[1,0]}$. We then check if this parametrisation is constrained further by the condition
\begin{equation}
\eta\wedge\kdel\db_{[1,1]}=\kbdel\rho_{[3,0]},\label{eq:constraint_1}
\end{equation}
which restricts to those $\eta\wedge\db_{[1,1]}$ for which a potential $\rho_{[3,0]}$ exists. Given such a $\rho_{[3,0]}$, the most general $\rho_{[3,0]}$ is a sum of these contributions plus a $\kbdel$-closed component, up to modding out by $\eta\wedge\kdel\lambda_{[1,0]}$, where $\eta\wedge\bar{\partial}_{b}\lambda_{[1,0]}=0$.

Let us begin. First note that any $\kbdel$-closed element $\eta\wedge\db_{[1,1]}$ will actually satisfy $\eta\wedge\kbdel\db_{[1,1]}=0$ (since $\dd\eta=0$). As $\eta$ is nowhere vanishing, this can be the case only if
\begin{equation}
\kbdel\db_{[1,1]}=\eta\wedge\mu_{[0,2]}
\end{equation}
for some $\mu_{[0,2]}$. As $\db_{[1,1]}$ is charge $k-3$, $\mu_{[0,2]}$ is charge $k-6$. As mentioned above, $H_{\bar{\partial}_{b}}^{[0,2]}(k')$ is trivial for $k'>-3$. As we are assuming $k\geq3/2$, there are values of the charge that have non-trivial $[0,2]$ classes, in particular they can be present for $3/2\leq k\leq3$. Let us denote a basis for $H_{\bar{\partial}_{b}}^{[0,2]}(k-6)$ as $h_{[0,2]}^{a}$ -- recall from Section \ref{sec:trans_laplace} that this basis is finite dimensional. One then has
\begin{equation}
\begin{split}\kbdel\db_{[1,1]} & =\eta\wedge(c_{a}h_{[0,2]}^{a}+\kbdel\mu_{[0,1]})\\
\Rightarrow\quad\kbdel(\db_{[1,1]}+\eta\wedge\mu_{[0,1]}) & =\eta\wedge c_{a}h_{[0,2]}^{a},
\end{split}
\label{eq:gamma_11}
\end{equation}
where $c_{a}\in\bC$ and the right-hand side is manifestly $\kbdel$-closed, charge-$(k-3)$ and type-$[1,2]$. The relevant cohomology for these objects is $H_{\bar{\partial}_{b}}^{[1,2]}(k-3)$, which by the relations \eqref{eq:KR-SE5-1} and \eqref{eq:KR-SE5-2} is given by 
\begin{equation}
\KRH{1}{2}{k-3} \simeq \KRH{2}{0}{3-k} \simeq  \TVH{2}{0}{3-k}\oplus  \TVH{1}{0}{3-k},\label{eq:cohom_decomp_1}
\end{equation}
and hence by \eqref{eq:TVsummary} vanishes for $k\geq3/2$.
Thus for each $\eta\wedge h_{[0,2]}^{a}$ we can choose a $\kbdel$-potential $\gamma_{[1,1]}^{a}$ such that $\eta\wedge h_{[0,2]}^{a}=\kbdel\gamma_{[1,1]}^{a}$, allowing us to rewrite \eqref{eq:gamma_11} as
\begin{equation}
0=\kbdel(\db_{[1,1]}+\eta\wedge\mu_{[0,1]}-c_{a}\gamma_{[1,1]}^{a}).
\end{equation}
Since $H_{\bar{\partial}_{b}}^{[1,1]}(k-3)$ can be non-trivial in general, we introduce a basis of forms $h_{[1,1]}^{i}$ (which is again finite dimensional) with coefficients $c_{i}\in\bC$. Integrating the previous relation then gives
\begin{equation}
c_{i}h_{[1,1]}^{i}+\kbdel\mu_{[1,0]}=\db_{[1,1]}+\eta\wedge\mu_{[0,1]}-c_{a}\gamma_{[1,1]}^{a},
\end{equation}
where $\mu_{[1,0]}$ accounts for any $\kbdel$-exact components. The most general solution to $\eta\wedge\kbdel\db_{[1,1]}$ is thus
\begin{equation}
\eta\wedge\db_{[1,1]}=c_{a}\,\eta\wedge\gamma_{[1,1]}^{a}+c_{i}\,\eta\wedge h_{[1,1]}^{i}+\eta\wedge\kbdel\mu_{[1,0]},\label{eq:b11_sol}
\end{equation}
where one can show that the terms on the right-hand side are linearly independent (so we are not over counting).

We now check to see if \eqref{eq:constraint_1} further constrains our parametrisation of $\db_{[1,1]}$. Taking $\kdel$ of $\eta\wedge\db_{[1,1]}$, we find
\begin{equation}
\begin{split}
\eta\wedge\kdel\db_{[1,1]} & =c_{a}\,\eta\wedge\kdel\gamma_{[1,1]}^{a}+c_{i}\,\eta\wedge\kdel h_{[1,1]}^{i}+\eta\wedge\kdel\kbdel\mu_{[1,0]} \\
& =\kbdel(c_{a}\gamma_{[3,0]}^{a}+c_{i}\gamma_{[3,0]}^{i}+\eta\wedge\kdel\mu_{[1,0]}),
\end{split}
\end{equation}
where, since $\eta\wedge\gamma_{[1,1]}^{a}$ and $\eta\wedge h_{[1,1]}^{i}$ are both $\kbdel$-closed and $H_{\bar{\partial}_{b}}^{[3,1]}(k)$ is trivial (as it is isomorphic to $ \TVH01k$), we have used
\begin{equation}
\eta\wedge\kdel\gamma_{[1,1]}^{a}=\kbdel\gamma_{[3,0]}^{a},\qquad\eta\wedge\kdel h_{[1,1]}^{i}=\kbdel\gamma_{[3,0]}^{i},
\end{equation}
for some $\gamma_{[3,0]}^{a}$ and $\gamma_{[3,0]}^{i}$. These degrees of freedom can always be used to solve \eqref{eq:constraint_1} without imposing any further conditions on $\db_{[1,1]}$, leaving only modding out with respect to $\kbdel$-closed $\eta\wedge\lambda_{[1,0]}$.

In summary, at this point we have that for $k\geq3/2$
\begin{equation}
\etaH 2k \simeq \KRH{0}{2}{k-6} \oplus \KRH{1}{1}{k-3} \oplus \tilden ,
\end{equation}
where $\tilden$ is the cohomology of the complex
\begin{equation}\label{eq:reduced_complex_2}
\begin{tikzcd}[column sep={4em,between origins}, row sep={4em,between origins}]
& \eta\wedge\lambda_{[1,0]} \arrow[rd, "\partial_b"] \arrow[ld, "\bar\partial_b"'] & & \\ 
0 & & \rho_{[3,0]} \arrow[rd, "\partial_b"] \arrow[ld, "\bar\partial_b"'] & \\
& 0 & & 0
\end{tikzcd}
\end{equation}Note that $\rho_{[3,0]}$ is automatically $\kdel$-closed (since it is type $[3,0]$) and so we need only constrain $\rho_{[3,0]}$ to be a $\kbdel$-closed charge-$k$ $[3,0]$-form. The choices of $\rho_{[3,0]}$ are thus counted by the dimension of $\KRH{3}{0}{k}$.

We then need to mod out by $\kbdel$-closed $\eta\wedge\lambda_{[1,0]}$. This implies that we have
\begin{equation}
\kbdel\lambda_{[1,0]}=\eta\wedge\mu_{[0,1]}\label{eq:db_closed}
\end{equation}
for some $\mu_{[0,1]}$, which in turn requires that $\mu_{[0,1]}$ is $\kbdel$-closed. As $\KRH{0}{1}{k-6}$ is trivial, $\mu_{[0,1]}$ must be $\kbdel$-exact, allowing us to write $\mu_{[0,1]}=\kbdel\alpha_{[0,0]}$ for some $\alpha_{[0,0]}$. Since $\eta\wedge\kbdel\alpha_{[0,0]}$ is in the kernel of ${}\eta\wedge$, without loss of generality we can take $\lambda_{[1,0]}$ to be $\kbdel$-closed. Considering the image of $\kdel(\eta\wedge\lambda_{[1,0]})$ in $\rho_{[3,0]}$ with $\lambda_{[1,0]}$ $\kbdel$-closed, one then has
\begin{equation}
\tilden \simeq \op{coker}(\eta\wedge\kdel),
\end{equation}
where $\eta\wedge\kdel$ maps from $\lambda_{[1,0]}\in \KRH{1}{0}{k-3}$ to $\rho_{[3,0]}\in \KRH{3}{0}{k}$. Since these are all finite-dimensional spaces, we can equally write this as
\begin{equation}
\dim\tilden=\dim \KRH{3}{0}{k}-\dim \KRH{1}{0}{k-3}+\dim\ker(\eta\wedge\kdel).
\end{equation}
Moreover, one can show that a $\kbdel$-closed, charge-$(k-3)$ $[1,0]$-form $\lambda_{[1,0]}$ which satisfies $\eta\wedge\kdel\lambda_{[1,0]}=0$ can always be written as
\begin{equation}
\lambda_{[1,0]}=h\,\eta+\kdel h',\label{eq:kernel_eta_del}
\end{equation}
where $h$ and $h'$ are holomorphic ($\bdel$- or $\kbdel$-closed) functions of charge $k-6$ and $k-3$ respectively. We can then translate the kernel of $\eta\wedge\kdel$ acting on $\KRH{0}{1}{k-3}$ into a statement about the image of a map $\kappa$ acting on these functions:
\begin{equation}
\begin{split}\kappa & \colon \KRH{0}{0}{k-6}\oplus \KRH{0}{0}{k-3}\to \KRH{1}{0}{k-3}\\
\kappa(h,h') & =h\,\eta+\kdel h'\equiv h\,\eta+\dd h'.
\end{split}
\end{equation}
The result of \eqref{eq:kernel_eta_del} is that $\ker(\eta\wedge\kdel)=\image\kappa.$ As the relevant spaces are again finite dimensional, we have
\begin{equation}
\dim\ker(\eta\wedge\kdel)=\dim \KRH{0}{0}{k-6}+\dim \KRH{0}{0}{k-3}-\dim\ker\kappa.
\end{equation}
We are left with calculating the dimension of the kernel of $\kappa$.

We begin with an observation: taking the exterior derivative of $\kappa(h,h')=0$ implies we must have $\dd(h\,\eta)=0$. Expanding out the derivative and using that $\eta$ is $\dd$-closed, one sees that the vanishing of $\dd(h\,\eta)$ also implies $\kbdel h=0$, and so $h$ is automatically holomorphic. Different choices of $h$ are thus counted by the ``degree-zero $\eta$-cohomology'' -- functions $h$ that satisfy $\eta\wedge\dd h=0$. It is relatively simple to show that a charge-$3t$ holomorphic function $h$ that satisfies this condition must take the form
\begin{equation}
h=\begin{cases}
cf^{t} & t\in\{0,1,2,\ldots\},\\
0 & \text{otherwise},
\end{cases}
\label{eq:h_three}
\end{equation}
where $c\in\bC$ and $f$ is the charge-$3$ holomorphic function that defines $\eta=\dd f$. Thus, the space of $\dd$-closed $h\,\eta$ is one dimensional whenever $t$ is a non-negative integer, and zero dimensional otherwise. One can then show that the kernel of $\kappa$ is generated by $(0,1)$ for $t=-1$ and $((t+1)f^{t},-f^{t+1})$ for $t=0,1,\ldots$, and vanishes otherwise. Given that $h$ has charge $k-6$, we can rewrite this condition in terms of $k$ as
\begin{equation}
\dim\ker\kappa=\begin{cases}
1 & k\equiv_{3}0,~k\geq3,\\
0 & \text{otherwise},
\end{cases}
\end{equation}
where $\equiv_{3}$ should be read as modulo 3.

In summary, we have shown, under the assumption that $k\geq 3/2$,
\begin{equation}
\begin{split}
\dim\etaH 2k & = \KRh{0}{2}{k-6} + \KRh{1}{1}{k-3} + \dim\tilden,\\
\dim\tilden & =\KRh{3}{0}{k}-\KRh{1}{0}{k-3}+\dim\ker(\eta\wedge\kdel),\\
\dim \ker(\eta\wedge\kdel) & =\KRh{0}{0}{k-6}+\KRh{0}{0}{k-3}-\dim\ker\kappa,\\
\dim\ker\kappa & =[k\equiv_{3}0,k\geq3],
\end{split}
\end{equation}
where  we are using ``Iverson brackets'' that evaluate to 1 if the contained statement is true, and 0 otherwise, and we have introduced the notation $\KRh pqk=\dim\KRH pqk$ for the dimensions of the cohomology groups. 

Using the dualities~\eqref{eq:TVsummary}, so that all terms have the same charge, we finally have
\begin{equation}
\begin{split}
\dim\etaH 2k &= \KRh{3}{2}{k-3}
    + \KRh{3}{0}{k-3} + \KRh{1}{1}{k-3} % \\ & \quad
    + 2\KRh{0}{0}{k-3} - \KRh{1}{0}{k-3}
    -[k\equiv_{3}0], \qquad \text{for $k\geq3/2$.} 
\end{split}
\end{equation}
Using the isomorphisms between the Kohn--Rossi and transverse Dolbeault cohomologies, writing $\TVh pqk$ for $\dim \TVH pqk$ and noting that $\TVh{0}{2}{k-3}$ and $\TVh{0}{1}{k-3}$ both vanish for $k\geq3/2$ (and $\TVh 01k$ and $\TVh 21k$ vanish identically), we can rewrite this in the more symmetric form 
\begin{equation}
\label{eq:etaH2}
\begin{split}
\dim\etaH 2k 
%     &= \TVh{2}{2}{k-3} +\TVh{1}{1}{k-3}
%     + \TVh{0}{0}{k-3} % \\ &\quad 
%     - \TVh{1}{0}{k-3} - \TVh{0}{1}{k-3}
%    + \TVh{2}{0}{k-3} % \\ &\quad 
%    + \TVh{0}{2}{k-3} - [k\equiv_{3}0] , \\
    &= \sum_{p,q=0,1,2} (-1)^{p+q} \TVh pq{k-3} - [k\equiv_{3}0] ,  
    \qquad \text{for $k\geq3/2$.} 
\end{split}
\end{equation}

In the next subsection, we will argue\footnote{We expect this can also be shown directly from the double complex using a $\partial\bar{\partial}$-type lemma for $\kdel$ and including representatives for the $\TVH20k$ and $\TVH01k$ groups that vanished for the case $k\geq3/2$.} that \eqref{eq:etaH2} actually holds for \emph{all $k$}. In this case, the relation $\TVh pq{-k}=\TVh {2-p}{2-q}k$ implies that $\etaH 2{k+3}\simeq \etaH 2{3-k}$ which together with the duality $\etaH2k=\etaH{2}{3-k}$, implies that $\etaH 2k$ is periodic in $k$: $\etaH 2k=\etaH 2{k+3}$. Hence, we can write 
\begin{equation}
\label{eq:etaH2-allk}
\begin{split}
\dim\etaH 2k 
%     &= \TVh{2}{2}{k-3} +\TVh{1}{1}{k-3}
%     + \TVh{0}{0}{k-3} % \\ &\quad 
%     - \TVh{1}{0}{k-3} - \TVh{0}{1}{k-3}
%    + \TVh{2}{0}{k-3} % \\ &\quad 
%    + \TVh{0}{2}{k-3} - [k\equiv_{3}0] , \\
    &= \sum_{p,q=0,1,2} (-1)^{p+q} \TVh pq{k} - [k\equiv_{3}0] , 
    \qquad \text{for all $k$,} \\
    &= \TVindex k - [k\equiv_{3}0] , 
\end{split}
\end{equation}
where in the second line $\TVindex k$ is the transverse Dolbeault cohomology index defined in Section \ref{sec:index}. Note that one consequence of this relation is that it implies the index is also periodic: $\TVindex{k+3}=\TVindex k$.

\subsection{\texorpdfstring{$\etaH 0k$}{H(d eta)(2)(k)}, \texorpdfstring{$\etaH 1k$}{H(d eta)(1)(k)} and the index of the \texorpdfstring{$\eta$}{eta}-complex}

Since we are in five dimensions, one can define the group $\etaH sk$ for $s=0,1,2,3,4$. The duality~\eqref{eq:charge_sym} states $\etaH sk\simeq\etaH {4-s}{-k}$ and furthermore $\etaH sk=0$ for $s>4$. Hence, determining $\etaH 0k$ and $\etaH 1k$, together with the results of the previous section, completes the calculation of the $\eta$-cohomologies.  

The degree-zero cohomology $\etaH0k$ simply counts the number of $\dd$-closed one-forms $\alpha\eta$, where $\alpha$ is a function. Recall that we have actually seen precisely this problem around \eqref{eq:h_three}. Using this previous result, we have
\begin{equation}
    \etaH0k = [k\equiv_{3}0,k\geq3] \bC.
\end{equation}
% Note that $n_{0,k}=0$ for $k<3$, so from (\ref{eq:duality}) we also have 
% \begin{equation}
% n_{4,k}=0\quad\text{for }k>0.
% \end{equation}

Next consider the degree-one cohomology $\etaH1k$. Again we can split into $[p,q]$ type to give
\begin{equation}\label{eq:reduced_complex_s_1}
\begin{tikzcd}[column sep={4em,between origins}, row sep={4em,between origins}]
&  & \eta\wedge\alpha_{[0,0]} \arrow[rd, "\partial_b"] \arrow[ld, "\bar\partial_b"'] & &  \\ 
& \eta\wedge \lambda_{[0,1]} \arrow[rd, "\partial_b"] \arrow[ld, "\bar\partial_b"'] & & \eta\wedge\lambda_{[1,0]} \arrow[rd, "\partial_b"] \arrow[ld, "\bar\partial_b"'] & \\
0 & & 0 & & 0
\end{tikzcd}
\end{equation}
and proceed as we did to calculate $\etaH2k$. First note that $\kbdel$-closure of $\eta\wedge\lambda_{[0,1]}$ implies $\kbdel\lambda_{[0,1]}=0$. Since $\KRH{0}{1}{k}\simeq H_{\kbdel}^{(0,1)}(k)$ and this is trivial (see (\ref{eq:vanishing_0q})), we have $\lambda_{[0,1]}=\kbdel\tilde{\alpha}_{[0,0]}$ for some $\tilde{\alpha}_{[0,0]}$, and so $\lambda_{[0,1]}$ can always be set to zero using the freedom to shift by $\kbdel(\eta\wedge\alpha_{[0,0]})$. The complex then reduces to
\begin{equation}\label{eq:reduced_complex_s_2}
\begin{tikzcd}[column sep={4em,between origins}, row sep={4em,between origins}]
& \eta\wedge\alpha_{[0,0]} \arrow[rd, "\partial_b"] \arrow[ld, "\bar\partial_b"'] & & \\ 
0 & & \eta\wedge\lambda_{[1,0]} \arrow[rd, "\partial_b"] \arrow[ld, "\bar\partial_b"'] & \\
& 0 & & 0
\end{tikzcd}
\end{equation}
where, as we saw around (\ref{eq:db_closed}), we can restrict to $\kbdel$-closed $\lambda_{[1,0]}$ and $\alpha_{[0,0]}$ without loss of generality. Using the results around (\ref{eq:kernel_eta_del}), $\lambda_{[1,0]}$ can always be written as $\lambda_{[1,0]}=h\,\eta+\kdel h',$where $h$ and $h'$ are holomorphic functions of charge $k-6$ and $k-3$ respectively. This means that $\eta\wedge\lambda_{[1,0]}$ can always be written as $\eta\wedge\lambda_{[1,0]}=\eta\wedge\kdel h'.$ However, all such elements are in the image of $\kdel$ acting on $\eta\wedge\alpha_{[0,0]}$ and so $\lambda_{[1,0]}$ can also be set to zero. This means that there are no non-trivial elements of the cohomology. This holds for all $k$ and so we have 
\begin{equation}
    \etaH1k = 0 . 
\end{equation}

In summary, using the dualities~\eqref{eq:charge_sym} we have
\begin{equation}
\label{eq:etaHother}
\begin{aligned}
    \etaH0k & = [k\equiv_{3}0,k\geq3] \bC, & &&
    \etaH1k & = 0, \\
    \etaH3k &= 0, & &&
    \etaH4k & = [k\equiv_{3}0,k\leq 0] \bC, 
\end{aligned}
\end{equation}
with $\etaH2k$ given by~\eqref{eq:etaH2-allk}. Although the $\eta$-complex is not elliptic, the cohomology groups are all finite dimensional and, as for the transverse Dolbeault operator, we can define an index
\begin{equation}
    \etaindex k \equiv \sum_n (-1)^n \dim \etaH nk . 
\end{equation} 
Substituting from \eqref{eq:etaHother} and \eqref{eq:etaHother}, we find that the $\eta$-complex index and  transverse Dolbeault index are equal:  
\begin{equation}
\label{eq:etaindex-TVindex}
    \etaindex k = \TVindex k . 
\end{equation}
As we will argue, in the dual field theory both expressions encode the single-trace superconformal index and hence should agree, since one theory is simply a marginal deformation of the other. This relation is true if \eqref{eq:etaH2} holds for all $k$ -- this is one motivation for making this assumption in the previous subsection. 

\section{Counting field theory operators}\label{sec:indices}

As we discussed in the introduction, AdS/CFT relates the reduced cyclic homology groups $\rHC_n(A,k)$ that count short multiplets of operators in the field theory to cohomology groups defined on $M$ that count certain Kaluza--Klein modes in the $\AdS5\times M$ type IIB supergravity background. In this section, we will first review how this works when $M$ is a Sasaki--Einstein space, following \cite{Eager:2012hx}. This gives a standard relation between $\rHC_n(A,k)$ and $\TVH pqk$, and hence an expression for the superconformal index in terms of transverse cohomologies. We then argue that the $\eta$-cohomology groups count the corresponding modes on the exceptional Sasaki--Einstein space that is dual to a finite exactly marginal deformation of the original theory, and hence give the cyclic homologies of the corresponding deformed Calabi--Yau algebra. Recall that throughout we use a normalisation where the conventional R-charge is given by $R=\frac23k$, meaning the superpotential has $\xi$-charge 3. 

\subsection{The undeformed theory}

Let us review the results of~\cite{Eager:2012hx}. Let $M$ be a five-dimensional Sasaki--Einstein manifold. By an explicit identification of the Kaluza--Klein modes, the authors of~\cite{Eager:2012hx} show that short supergravity multiplets are counted by the transverse cohomology groups\footnote{Recall that~\cite{Eager:2012hx} refers  to $\TVH pqk$ (rather than $\KRH pqk$) as the ``Kohn--Rossi cohomology groups''.} $\TVH pqk$ and furthermore identify the form of the operators in the SCFT to which each mode is dual. Following their notation and taking $k>0$, one has 
\begin{equation}
\label{eq:SE-operators}
\begin{aligned}
   \TVH 00k\colon & \qquad 
    \tr \mathcal{O}_f, \ \tr W_\alpha\mathcal{O}_f , \
    \tr W_\alpha W^\alpha\mathcal{O}_f , && && +t^{2k}, \\
    \TVH 11k\colon & \qquad 
    \tr \mathcal{O}_\omega, \ \tr W_\alpha\mathcal{O}_\omega , \
    \tr W_\alpha W^\alpha\mathcal{O}_\omega , && && +t^{2k}, \\
    \TVH 10k\colon & \qquad 
    \tr \mathcal{O}_v, \ \tr W_\alpha\mathcal{O}_v , \
    \tr W_\alpha W^\alpha\mathcal{O}_v , && && - t^{2k}, \\
    \TVH 20k\colon & \qquad
    \tr \bar{W}_{\dot{\alpha}}\mathcal{O}_{f'}, \ 
    \tr \bar{W}_{\dot{\alpha}}W_\alpha\mathcal{O}_{f'} , \
    \tr \bar{W}_{\dot{\alpha}}W_\alpha W^\alpha\mathcal{O}_{f'} , && && + t^{2k}, 
\end{aligned}
\end{equation}
where the labels $f$, $v$ and $\omega$ are charge-$k$ elements of $\TVH 00k$, $\TVH 10k$ and $\TVH11k$ respectively, while the function $f'$ is a charge-$(k-3)$ element of $\TVH 00{k-3}\simeq\TVH 20k$. The dual supergravity modes are constructed explicitly in terms of $f$, $v$, $\omega$ and $f'$. The final term in each line of~\eqref{eq:SE-operators} is the net contribution of the three operators to the single-trace superconformal index $\Ist(t)$ of the SCFT~\cite{Romelsberger:2005eg,Kinney:2005ej}.

The operators in the first two lines of \eqref{eq:SE-operators} are of the same type: scalar chiral, spinor chiral and scalar chiral as one reads across. In the field theory their contribution to the index is collectively counted by $\rHC_0(A,k)$. The contribution of the operators in the third and fourth lines on the other hand are counted by $\rHC_1(A,k)$ and $\rHC_2(A,k)$ respectively. We see that the AdS/CFT correspondence hence predicts the relation
\begin{equation}
\label{eq:HC-TVH}
    \rHC_n(A,k) \simeq \bigoplus_{p-q=n} \TVH pqk \, [k>0]
\end{equation} 
where we have used the fact that $\TVH22k$ and $\TVH21k$ vanish for $k>0$. For a Sasaki--Einstein manifold $M$, the algebra $A$ has the same cyclic homology as the coordinate ring of the cone over $M$, and one can show directly that the relation \eqref{eq:HC-TVH} indeed holds \cite{Eager:2012hx}. Calculating the single-trace superconformal index gives
\begin{equation}
\label{eq:SEindex}
\begin{split}
    \Ist(t) &= \sum_{0\leq n \leq 2,\ k>0} (-1)^n t^{2k}\dim\rHC_n(A,k) \\
       &= \sum_{0\leq p,q\leq 2,\ k>0} (-1)^{p-q} t^{2k} \dim \TVH pqk 
       = \sum_{k>0} t^{2k} \TVindex k .
\end{split}
\end{equation}

Rather than performing a full Kaluza--Klein analysis, one can also count multiplets by considering supersymmetric perturbations of the background following~\cite{Ashmore:2016oug}. Solving for a linear deformation of the geometry that preserves part of the integrability of the hypermultiplet structure (H-structure) defined in~\cite{Ashmore:2015joa,Ashmore:2016qvs}, modulo diffeomorphisms and gauge transformations, identifies the perturbation with elements of $\TVH00k$ and $\TVH11k$. Such deformations are dual to a scalar chiral primary operator. In particular, they either correspond to perturbing the SCFT by the $F$-term of a scalar chiral operator $\mathcal{C}=A+\theta\psi+\theta^2F$ or to giving a vacuum expectation value (vev) to the lowest component $A^*$ of the anti-chiral operator $\bar{\mathcal{C}}=A^*+\bar{\theta}\bar{\psi}+\bar{\theta}^2F^*$. Focussing on the former, if $n_k(\mathcal{C})$ is the number of $F$-term deformations of charge $k$, one finds 
\begin{equation}
\label{eq:chiral-ops}
    \begin{aligned}
        n_k(\tr\mathcal{O})&= \bar{n}^0_k , & &&
        n_k(\tr W_\alpha W^\alpha\mathcal{O}) &= \bar{n}^0_{k-3} + (b_2+1) [k=3] , 
    \end{aligned}
\end{equation}
where $\mathcal{O}$ is $\mathcal{O}_f$ or $\mathcal{O}_w$ and we have defined
\begin{equation}
\label{eq:nbar-def}
    \bar{n}^0_k \equiv \dim \rHC_0(A,k) = (\TVh 00k+\TVh 11k)[k>0] ,
\end{equation}
and $b_2\equiv\dim H_\dd^2(M)=\TVh 110-1$ is the second Betti number. Note that for charge-zero $f$ or $w$, there is no corresponding $F$-term deformation of the form $\tr\mathcal{O}_f$ and $\tr\mathcal{O}_w$. This is because the bulk supergravity modes are dual to $\SU N$ rather than $\Uni{N}$ quiver gauge theories and so there are no operators of the form ``$\tr\mathbbm{1}$''. However, the corresponding terms are present for the $\tr W_\alpha W^\alpha\mathcal{O}_f$ and  $\tr W_\alpha W^\alpha\mathcal{O}_w$ operators at $k=3$. They give marginal perturbations of the overall and relative coupling constants associated to different gauge groups in the quiver respectively. We will write these operators heuristically as $\tr W_\alpha W^\alpha$ and $\tr W_\alpha W^\alpha-\tr W'_\alpha W'^\alpha$. They are dual respectively to constant perturbations of the axion-dilaton and to turning on a non-trivial complex two-form potential (with vanishing flux) in the type IIB supergravity, and hence are counted by $b_2+1$. In the notation of \cite{Ashmore:2016oug}, the $\mathcal{O}_f$ and  $\tr W_\alpha W^\alpha\mathcal{O}_f$ $F$-terms are dual to the modes labelled by $f$ with $k>0$ and $\bar{f}$ with $k\geq0$ respectively,\footnote{\label{fn:S5} There is a subtlety, relevant to the S$^5$ theory, that the supergravity analysis excludes any $k=1$ mode for $f$. Thus the number of modes is not given by $\bar{n}^0_1=3$. This is however completely consistent with the fact that $\tr\Phi^i=0$ for the $\SU{n}$ theory.} while the $\mathcal{O}_w$ and  $\tr W_\alpha W^\alpha\mathcal{O}_w$ $F$-terms are dual to the modes labelled by $\delta'$ with $k>0$ and $\delta$ with $k\geq0$ respectively.

Writing $n_k(\overline{C})$ for the number of supersymmetric vevs with charge $k$, one also finds 
\begin{equation}
\label{eq:anitchiral-ops}
    \begin{aligned}
        n_k(\tr\overline{\mathcal{O}})&= \bar{n}^0_{3-k} , & &&
        n_k(\tr \overline{W_\alpha W^\alpha\mathcal{O}}) &= \bar{n}^0_{-k} + (b_2+1) [k=0] , 
    \end{aligned}
\end{equation}
where again $\mathcal{O}$ is $\mathcal{O}_f$ or $\mathcal{O}_w$. In the notation of \cite{Ashmore:2016oug}, the $\tr\overline{\mathcal{O}_f}$ and  $\tr\overline{W_\alpha W^\alpha\mathcal{O}_f}$ vevs are dual to the modes labelled by $(f')^*$ with $k<0$ and $(\bar{f}')^*$ with $k\leq0$ respectively (where here the star denotes complex conjugation), while the $\tr\overline{\mathcal{O}_w}$ and  $\tr\overline{W_\alpha W^\alpha\mathcal{O}_w}$ vevs are dual to the modes labelled by $\delta$ with $k<0$ and $\delta'$ with $k\leq0$ respectively.

Putting everything together, we can write an expression for the total number $m_k$ of supersymmetric perturbations of charge $k$, dual to both deformations and vevs as 
\begin{equation}
\label{eq:mk-def}
    m_k = q^0_k + q^0_{3-k},
\end{equation}
where 
\begin{equation}
\label{eq:q0k-SE}
    q^0_k = \bar{n}^0_k + \bar{n}^0_{-k} + (b_2+1)[k=0] ,
\end{equation}
so that $q^0_{-k}=q^0_k$.

\subsection{The deformed theory: chiral multiplets and \texorpdfstring{$\etaH 2k$}{H(d eta)(2)(k)}}

We now turn to the counting of short operators for the marginally deformed theories. As we have discussed in Section~\ref{sec:eta-complex}, in this case the H-structure is (partly) determined by the CR structure of the original Sasaki--Einstein manifold together with a one-form $\eta$ of charge three. Furthermore, the supersymmetric deformations are counted by the $\eta$-cohomology~\cite{letter}. 

Explicitly one finds that the total number of scalar chiral perturbations, including by deformations and vevs, is given by 
\begin{equation}
\label{eq:mk-deform}
\begin{aligned}
    m_k &= 2 \dim \etaH 2k - [k=0] - [k=3]  \\
        &= \bigl(\dim \etaH 2k - [k=0]\bigr) 
           + \bigl(\dim \etaH 2{3-k} - [k=3]\bigr) ,
\end{aligned}
\end{equation}
where we have used the duality $\etaH 2k\simeq\etaH 2{3-k}$. 

Using the relation \eqref{eq:mk-def}, we would like to find $q^0_k$ and hence relate the $\eta$-cohomology $\etaH 2k$ to the reduced cyclic homology $\rHC_0(A,k)$ for the deformed field theory. However, as it stands\footnote{In \cite{letter}, the supersymmetric deformations were counted using only a particular equivalence class of the H-structure. Giving an explicit form for the full geometry remains an open problem. If this were known one could, in analogy to \cite{Ashmore:2016oug}, identify the representative corresponding to each mode of $q^0_k$ and $q^0_{k-3}$ separately.} we cannot unambiguously read off $q^0_k$ from~\eqref{eq:mk-deform} unless one knows the expression for $q^0_k$ for $0\leq k\leq \frac{3}{2}$. However, we can use a simple physical argument to solve this problem as follows. For $0< k\leq \frac32$, the parameters $q^0_k$ count $F$-term deformations by operators of the form $\tr\mathcal{O}$, all of which are relevant in this range of $k$. For $k=0$, the parameter $q^0_0$ counts the vevs of the operators of the form $\tr\overline{W_\alpha W^\alpha}$ and $\tr\overline{W_\alpha W^\alpha}-\tr\overline{W'_\alpha W'^\alpha}$. As we have discussed, the corresponding $\bar{F}$-term components give marginal deformations that deform the coupling constants of the gauge theory. From the analysis of \cite{Leigh:1995ep,Kol02,GKSTW10}, all marginal operators of this form are exactly marginal. Furthermore, the number of relevant (or exactly marginal) operators cannot change under a finite marginal deformation of the field theory. Thus, in this window $0\leq k\leq\frac32$ we expect that $q^0_k$ is the same as in the \emph{undeformed} theory.\footnote{This relation should actually hold for the whole range $0\leq k<3$, since all the operators in $\frac32<k<3$ are relevant. We will see this is indeed so in the examples we consider below, although care must be taken in the S$^5$ case due to the subtlety noted in Footnote \ref{fn:S5}} From~\eqref{eq:etaH2-allk} we note that, for $0\leq k \leq\frac32$,
\begin{equation}
\label{eq:q0k_window}
    \dim \etaH 2{k} - [k=0] = (\TVh00k + \TVh11k)[k>0] + (b_2+1)[k=0], 
\end{equation}
where we have used the relations and bounds given in \eqref{eq:TVsummary}, and that $\TVh000=1$ and $\TVh110=b_2+1$. However, from \eqref{eq:nbar-def} and \eqref{eq:q0k-SE}, this is exactly equal to the undeformed $q^0_k$ in this window. Thus comparing \eqref{eq:mk-def} and \eqref{eq:mk-deform}, we find
\begin{equation}
    q^0_k = \dim \etaH 2k - [k=0] , 
\end{equation}
valid for all $k$. From \eqref{eq:q0k-SE} we finally have, for the deformed theory, 
\begin{equation}
\label{eq:HC-Heta}
    \rHC_0(A,k) \simeq \etaH 2{k}\,[k>0] ,
\end{equation}
relating the reduced cyclic homology and the $\eta$-cohomology. We can write down a \emph{Hilbert series} $\tilde{H}(t)$ that is just the generating function for the number of (single-trace) chiral operators of charge $k$. We define\footnote{In defining $\tilde{H}(t)$ we use the same power of twice the conformal dimension $t^{2k}$ that appears in the index. As we will see in the examples, when the R-symmetry is compact, this normalisation does not necessarily match the usual definition of the Hilbert series, where one normalises by the minimal $\Uni 1$ charge.}
\begin{equation}
\label{eq:Hilbert-def}
    \tilde{H}(t) \equiv 1 + \sum_{k>0} t^{2k} \dim \rHC_0(A,k) .
\end{equation}
From \eqref{eq:HC-Heta} and \eqref{eq:etaH2-allk} we see that, for all the deformed theories, the Hilbert series is related to the single-trace superconformal index by
\begin{equation}
\label{eq:Hilbert-index}
    \tilde{H}(t) = 1 + \Ist(t) - \frac{t^6}{1-t^6},
\end{equation}
where the expansion of $t^6/(1-t^6)$ around $t=0$ gives $[k\equiv_3,k>0]t^{2k}$.

\subsection{The deformed theory: semi-long multiplets and  \texorpdfstring{$\etaH 0k$}{H(d eta)(0)(k)} and  \texorpdfstring{$\etaH 1k$}{H(d eta)(1)(k)}}

As we have seen, $\dim\rHC_1(A,k)$ and $\dim\rHC_2(A,k)$ count the number of operators in the third and fourth lines of \eqref{eq:SE-operators}, all of which are short supersymmetric multiplets. For example, we note that $\tr W_\alpha\mathcal{O}_v$ is a spin-$(\frac12,0)$ semi-long multiplet, while $\tr \bar{W}_{\dot{\alpha}}\mathcal{O}_{f'}$ and $\tr \bar{W}_{\dot{\alpha}}W_\alpha W^\alpha\mathcal{O}_{f'}$ are spin-$(0,\frac12)$ semi-long multiplets. They have highest components that are vectors and two-forms respectively, and are also the only operators in the third and fourth lines of \eqref{eq:SE-operators} that have bosonic highest components. Thus, one should be able to deform the theory by turning on these vector or two-form operators and still preserve supersymmetry, albeit while breaking the Lorentz symmetry. Without working out the details, these are naturally related to deformations of the ``V-structure'' of \cite{Ashmore:2016qvs} in the dual supergravity, and should be counted by the $\etaH1k$ and $\etaH0k$ cohomologies respectively. Thus, combining with \eqref{eq:HC-Heta}, we are led to the general conjecture
\begin{equation}
\label{eq:HC-Heta-gen}
    \rHC_n(A,k) \simeq \etaH {2-n}k \, [k>0] . 
\end{equation}
In particular, we predict
\begin{equation}
\begin{aligned}
    \rHC_1(A,k) &= 0 , & &&
    \rHC_2(A,k) &= [k\equiv_3 0, k>0] \bbC . 
\end{aligned}
\end{equation}
Notably, this implies there are \emph{no operators} of the form $\tr W_\alpha\mathcal{O}_v$ (and hence also of the form $\tr \mathcal{O}_v$ and $\tr W_\alpha W^\alpha\mathcal{O}_v$) in the deformed theory.

The single-trace superconformal index for the Sasaki--Einstein and the deformed theory should of course be the same. From \eqref{eq:HC-Heta-gen}, we find
\begin{equation}
\begin{split}
    \Ist(k) &= \sum_{0\leq n \leq 2,\ k>0} (-1)^n t^{2k}\dim\etaH {2-n}k \\
       &= \sum_{k>0} t^{2k} \etaindex k .
\end{split}
\end{equation}
Using \eqref{eq:etaindex-TVindex}, we see that this indeed agrees with the index for the undeformed theory \eqref{eq:SEindex}.

\section{Examples}\label{sec:examples}

Let us now briefly specialise our results to a few simple examples of deformed theories. This will allow us to check the relation $\rHC_n(A,k)\simeq\etaH{2-n}k$ in some instances and in others make some predictions about the field theory. 

For simplicity, we will take the Sasaki--Einstein geometry $M$ to be regular, though, as we have noted, only quasi-regularity is required for our analysis. For regular geometries, $M$ is an S$^1$ fibration over a Kähler--Einstein base $B$. This implies $B$ is Fano and is one of $\mathbb{P}^2$, $\mathbb{P}^1\times\mathbb{P}^1$ or $\dP n$ for $3\leq n\leq 8$, where the del Pezzo surface $\dP n$ is $\mathbb{P}^2$ blown up at $n$ points \cite{Tian87,TianYau87}. The Dolbeault cohomology groups can then be calculated from the bundle-valued sheaf cohomologies on $B$ of the S$^1$ fibration. More precisely, one has 
\begin{equation}
\TVH pqk % \simeq H^{p,q}\bigl(B,\mathcal{O}(s)\bigr)
\simeq H^{q}\Bigl(B,\Omega^{p,0}(B)\otimes K_B^{-k/3}\Bigr), \qquad \text{with $\tfrac{1}{3}kI_B \in \bbZ_{\geq0}$} ,
\end{equation}
which are the standard Dolbeault cohomologies valued in the tensor product of a power of the anti-canonical bundle $K_B^{-1}$ with the holomorphic cotangent bundle of $B$. Here $I_B$ is the Fano index of $B$, i.e.~the largest positive integer such that $c_1(K_B^{1/I_B})$ is an integral class on $B$. Recall that the deformation is defined by the one-form $\eta=\dd f$, where $f$ is holomorphic and has charge three. Reducing to the base we can hence view $f$ as a section $s$ of the anti-canonical bundle $K_B^{-1}$. The requirement that $\eta$ is nowhere vanishing implies that there are no points where $s$ and $\del s$ both vanish. Equivalently it means that the divisor defined by $s=0$ is smooth. Except for $\dP8$, the linear system defined by $K_B^{-1}$ is fixed-point free and so, for smooth $B$, the divisor is indeed smooth for generic $f$ by Bertini's theorem. 

% ***
% Some Macaulay 2 code
% P = Proj (QQ[x,y,z])
% H = (s, p, q) -> rank (HH^q( cotangentSheaf (p, P)** OO_P (s)))
% S = (p, q) -> (( -10..10) / (s -> H(s, p, q)))
% S(0,0)
% S(1,0)
% S(1,1)
% S(0,1)
% S(0,2)
% S(2,0)
% S(2,1)
% S(1,2)
% S(2,2)

% P = Proj (QQ[x,y,z,w]/(x^2 + y^2 + z^2 + w ^2))
% H = (s, p, q) -> rank (HH^q( cotangentSheaf (p, P)** OO_P (s)))
% S = (p, q) -> (( -10..10) / (s -> H(s, p, q)))
% S(0,0)
% S(1,0)
% S(1,1)
% S(0,1)
% S(0,2)
% S(2,0)
% S(2,1)
% S(1,2)
% S(2,2)

% P = Proj (QQ[x,y,z,w]/(x^3 + y^3 + z^3 + w^3 - x*y*z - 10*w*z*y))
% H = (s, p, q) -> rank (HH^q( cotangentSheaf (p, P)** OO_P (s)))
% S = (p, q) -> (( -10..10) / (s -> H(s, p, q)))
% S(0,0)
% S(1,0)
% S(1,1)
% S(0,1)
% S(0,2)
% S(2,0)
% S(2,1)
% S(1,2)
% S(2,2)
% ***

\subsection{\texorpdfstring{S$^{5}$}{S5}}

For S$^5$, the base is $\mathbb{P}^2$ and $I_B=3$. Using the sheaf cohomologies and the various dualities from Section \ref{KR_sec}, one finds the independent transverse cohomologies for $\text{S}^{5}$ are given by
\begin{equation}
\begin{split}
  \TVh00k & = \tfrac{1}{2}(k+1)(k+2)\,[k\geq0] , \\
    \TVh10k & = (k-1)(k+1) \, [k\geq2] , \\
    \TVh11k & =[k=0],
    \end{split}
\end{equation}
and should be understood to be non-zero only for integer values of $k$. Since the minimal charge is $k=1 $, it is natural to write the Hilbert series \eqref{eq:Hilbert-def} as $\tilde{H}(t)=H(t^2)$, where, from \eqref{eq:HC-TVH}, for the undeformed theory we have the standard result 
\begin{equation}
    H(t) = 1 + 3t + 6t^2 + 10t^3 + \ldots = \frac{1}{(1-t)^3}.
\end{equation}
We also have the single trace index
\begin{equation}
    \Ist(t) = \frac{3t^2}{1-t^2} .
\end{equation}

As discussed in the introduction, the generic deformed theory has a superpotential of the form \cite{Leigh:1995ep}
\begin{equation}
\label{eq:WdefS5}
\begin{split}
    \mathcal{W} &= h \tr\bigl(\Phi^1\Phi^2\Phi^3-\Phi^3\Phi^2\Phi^1\bigr)
        \\ & \qquad \qquad 
        + f_\beta \tr\bigl(\Phi^1\Phi^2\Phi^3+\Phi^3\Phi^2\Phi^1\bigr) 
        + f_\lambda \tr\bigl( (\Phi^1)^3 + (\Phi^2)^3 + (\Phi^3)^3 \bigr) ,
\end{split}
\end{equation}
where the undeformed theory has $f_\beta=f_\lambda=0$. This gives an algebra $A$ that is just the polynomial ring on $\bbC^3$ where we associate $\Phi^i$ with the coordinates $(x,y,z)$. Recall that the function $f$ in $\eta=\dd f$ can be then be read off from the deformation part of \eqref{eq:WdefS5} as 
\begin{equation}
    f = 2f_\beta \,xyz + f_\lambda \big( x^3 + y^3 + z^3 \big) , 
\end{equation}
where we are restricting from the $\bbC^3$ cone to the sphere S$^5$. Counting the chiral operators from the dual deformed geometry, we have, from \eqref{eq:etaH2-allk}, 
\begin{equation}
    \dim \etaH 2k = \TVindex k - [k\equiv_3 0]
        = 3 [k\in\bbZ] - [k\equiv_3 0] ,
\end{equation}
so that the Hilbert series \eqref{eq:Hilbert-def} is given by $\tilde{H}(t)=H(t^2)$ where
\begin{equation}
\begin{split}
    H(t) % &=\sum_k n^0_k t^k = \sum_k \bigl(\dim \etaH 2{k+3} - [k=0]\bigr) t^k \\
    &= 1 + 3t + 3t^2 + 2t^3 + 3t^4 + 3t^5 + 2t^6+\ldots
    =\frac{(1+t)^3}{1-t^3}.
    \end{split}
\end{equation}
In addition, we have the general result $\etaH 1k=0$ and $\etaH 0k=[k\equiv_30]\bbC$. 

\begin{table}[t]
\centering
\renewcommand*{\arraystretch}{1.1}
\begin{tabular}{@{}l>{\raggedleft}p{1.2em}>{\raggedleft}p{1.2em}>{\raggedleft}p{1.2em}>{\raggedleft}p{1.2em}>{\raggedleft}p{1.2em}>{\raggedleft}p{1.2em}>{\raggedleft}p{1.2em}>{\raggedleft}p{1.2em}>{\raggedleft}p{1.2em}>{\raggedleft}p{1.2em}>{\raggedleft}p{1.2em}>{\raggedleft}p{1.2em}@{}}
\toprule
$k$ & $-4$ & $-3$ & $-2$ & $-1$ & $0$ & $1$ & $2$ & $3$ & $4$ &  $5$ & $6$ & $7$ \tabularnewline
\midrule
$\tr(\mathcal{O})$ &  &  &  &  &  & & $6$ & $2$ & $3$ & $3$ & $2$ & $3$ \tabularnewline
$\tr(W^{2}\mathcal{O})$ &   &  &  &  &  &  &  & $1$ & $3$ & $3$ & $2$ & $2$ \tabularnewline
$\tr(\overline{\mathcal{O}})$ & $3$ & $2$ & $3$ & $3$ & $2$ & $6$ & \tabularnewline
$\tr(\overline{W^{2}\mathcal{O}})$ & $3$ & $2$ & $3$ & $3$ & $1$ \tabularnewline
\midrule
Total & $6$ & $4$ & $6$ & $6$ & $3$ & $6$ & $6$ & $3$ & $6$ & $6$ & $4$ & $6$ \tabularnewline
\bottomrule
\end{tabular}
\caption{Counting of operators in the field theory dual to S$^{5}$ graded by $k$. % There are four towers of operators corresponding to those in \eqref{eq:SE-operators}, namely those of the form $\tr(\mathcal O)$, $\tr(W_\alpha W^\alpha \mathcal{O})$, and their conjugates. 
Note that the number of $\tr(W_\alpha W^\alpha \mathcal{O})$ operators, except for the case of $k=0,1,2$, is the same as the number of $\tr(\mathcal{O})$ operators after a shift of $k$ by 3.}
\label{tab:S5}
\end{table}

One can check that these predictions agree with the field theory analysis for a generic deformed superpotential \eqref{eq:WdefS5} by repeating the analysis in \cite{Berenstein:2000ux} (which was carried out for the case of the beta deformation where $f_\lambda=0$). However, the Sklyanin-type non-commutative algebra defined by \eqref{eq:WdefS5} is one of the prototypical examples of a Calabi--Yau algebra, and the reduced cyclic homology groups have actually already been calculated by Van den Bergh in~\cite{VdB92}. One finds that  $\rHC_n(A,k)$ is indeed isomorphic to $\etaH {2-n}k$ for $k>0$. The explicit counting of scalar chiral perturbations due to deformations and vevs is given in Table \ref{tab:S5}. Note that there is a subtlety in the counting of $\tr(\mathcal{O})$ for small $k$. As already mentioned, there are no charge-zero operators of the form $\tr\mathbbm{1}$ or charge-one operators of the form $\tr\Phi^i$ because we are in the $\SU n$ theory. Extremising the superpotential we get 
\begin{equation}
\label{eq:dW}
    C_i - \tfrac{1}{N}\mathbbm{1}\tr C_i = 0 ,
\end{equation}
where
\begin{equation}
    C_1 = (h+f_\beta)\Phi^2\Phi^3 + (h-f_\beta)\Phi^3\Phi^2 + 3f_\lambda (\Phi^1)^2 , \qquad \text{etc.}
\end{equation}
Note that the second term in \eqref{eq:dW} means that there is no constraint on $\tr C_i$, and hence we have six distinct operators of charge-two, just as in the undeformed theory (see \cite{Eager:2015hwa,Freedman:2005cg,Madhu:2007ew}). Thus, as expected, the counting of relevant operators is indeed unchanged under a marginal deformation. Although this counting disagrees with $\HC_0(A,k)$, the total counting does agree with \eqref{eq:mk-deform} and \eqref{eq:HC-Heta}. Note also that, for the $\Uni{n}$ theory, by contrast, there are three operators of charge-one, and three of charge-two once the constraints $C_i=0$ are accounted for, agreeing with $\dim\HC_0(A,k)$.

\subsection{\texorpdfstring{T$^{1,1}$}{T11}}

For $\mathrm{T}^{1,1}$ the base is $\mathbb{P}^1\times\mathbb{P}^1$ and $I_B=2$. Using the various dualities, one finds the independent transverse Dolbeault cohomologies are 
\begin{equation}
\begin{split}
  \TVh00k & = (s+1)^{2}\, [s\geq0], \\
  \TVh10k & = 2(s+1)(s-1)\, [s\geq2] ,\\
  \TVh11k & = 2\, [s=0], 
    \end{split}
\end{equation}
where $k=3s/2$ and $s$ takes integer values. Thus it is natural to write the Hilbert series as $\tilde{H}(t)=H(t^3)$. For the undeformed case we find the standard result
\begin{equation}
    H(t) = 1 + 4t + 9t^2+ 16t^3 + \ldots = \frac{1+t}{(1-t)^3} , 
\end{equation}
and the single trace index is 
\begin{equation}
    \Ist(t) = \frac{4t^3}{1-t^3} .
\end{equation}

The generic deformed theory has a superpotential of the form \cite{Benvenuti:2005wi}
\begin{equation}
\label{eq:WdefT11}
\begin{split}
    \mathcal{W} &= h \tr\big(A_1 B_{\dot1}A_2B_{\dot2}
           - A_1 B_{\dot2}A_2B_{\dot1}\big) 
        + f_\beta \tr\big(A_1 B_{\dot1}A_2B_{\dot2}
           + A_1 B_{\dot2}A_2B_{\dot1}\big) 
        \\ & \qquad 
        + f_2 \tr\big(A_1B_{\dot1}A_1B_{\dot1}
           + A_2 B_{\dot2}A_2B_{\dot2}\big)
        + f_3\tr\big(A_1 B_{\dot2}A_1B_{\dot2}
           + A_2 B_{\dot1}A_2B_{\dot1}\big) .
\end{split}
\end{equation}
The undeformed theory has $f_\beta=f_2=f_3$ and gives $A$ as the polynomial ring algebra on the conifold $\mathcal{C}\subset\bbC^4$ given by $z_1^2+z_2^2+z_3^2+z_4^4=0$, where one associates 
\begin{equation}
    \begin{pmatrix}
       z_3+\ii z_4 & z_1-\ii z_2 \\     
       z_1+\ii z_2 & -z_3+\ii z_4    
    \end{pmatrix}
    \quad \leftrightarrow \quad
    \begin{pmatrix}
       A_1B_{\dot1} & A_1B_{\dot2} \\     
       A_2B_{\dot1} & A_2B_{\dot2}    
    \end{pmatrix} .
\end{equation}
The function $f$ in $\eta=\dd f$ can be then be read off from the deformation part of \eqref{eq:WdefT11} as 
\begin{equation}
    f = f_\beta \big( z_1^2 + z_2^2 - z_3^2 - z_4^2 \big)
       + 2f_2 \big( z_3^2 - z_4^2 \big)
       + 2f_3 \big( z_1^2 - z_2^2 \big),
\end{equation}
where we are restricting from the conifold cone to the T$^{1,1}$ link. For the generic marginally deformed theory counting the chiral operators gives, from \eqref{eq:etaH2-allk}, 
\begin{equation}
    \dim \etaH 2k = \TVindex k - [k\equiv_3 0]
        = 4 [2k \equiv_3 0] - [k\equiv_3 0] ,
\end{equation}
so that the Hilbert series \eqref{eq:Hilbert-def} is given by $\tilde{H}(t)=H(t^3)$ with 
\begin{equation}
\begin{split}
    H(t) % &=\sum_k q^0_k t^k = \sum_k \bigl(\dim \etaH 2{k+3} - [k=0]\bigr) t^k \\
    &= 1 + 4t + 3t^2 + 4t^3 + 3t^4 + 4t^5 + 3t^6 + \ldots
    = \frac{1+4t+2t^2}{1-t^2} .
\end{split}
\end{equation}
In addition, we have again the general results $\etaH 1k=0$ and $\etaH 0k=[k\equiv_30]\bbC$. 

We have checked that this result is in agreement with an explicit counting of gauge-invariant chiral fields modulo the $F$-term relations of the deformed superpotential up to $k=21/2$. We could not find any direct calculation of the dimension of the cyclic homology of the non-commutative Calabi--Yau algebra $A$ defined by the deformed superpotential for $\text{T}^{1,1}$, and so this can be regarded as a prediction for the form of $\rHC_n(A,k)$. The complete counting of scalar chiral perturbations due to deformations and vevs is given in Table \ref{tab:T11}.

\begin{table}[t]
\centering
\renewcommand*{\arraystretch}{1.1}
\begin{tabular}{@{}l>{\raggedleft}p{1.2em}>{\raggedleft}p{1.2em}>{\raggedleft}p{1.2em}>{\raggedleft}p{1.2em}>{\raggedleft}p{1.2em}>{\raggedleft}p{1.2em}>{\raggedleft}p{1.2em}>{\raggedleft}p{1.2em}>{\raggedleft}p{1.2em}>{\raggedleft}p{1.2em}>{\raggedleft}p{1.2em}@{}}
\toprule
$k$ & $-6$ & $-\frac92$ & $-3$ & $-\frac32$ & $0$ & $\frac32$ & $3$ & $\frac92$ & $6$ & $\frac{15}2$ & $9$\tabularnewline
\midrule
$\tr(\mathcal{O})$ & &  &  &  & & $4$ & $3$ & $4$ & $3$ & $4$ & $3$ \tabularnewline
$\tr(W^{2}\mathcal{O})$ &  &  &  &  &  & & $2$ & $4$ & $3$ & $4$ & $3$ \tabularnewline
$\tr(\overline{\mathcal{O}})$ & $3$ & $4$ & $3$ & $4$ & $3$ & $4$ \tabularnewline
$\tr(\overline{W^{2}\mathcal{O}})$ & $3$ & $4$ & $3$ & $4$ & $2$ \tabularnewline
\midrule
Total & $6$ & $8$ & $6$ & $8$ & $5$ & $8$ & $5$ & $8$ & $6$ & $8$ & $6$ \tabularnewline
\bottomrule
\end{tabular}
\caption{Counting of operators in field theory dual to T$^{1,1}$ graded by $k$. 
Note that the number of $\tr(W_\alpha W^\alpha \mathcal{O})$ operators, except for the case of $k=0$, is the same as the number of $\tr(\mathcal{O})$ operators after a shift of $k$ by 3. There are two modes for $\tr(W_\alpha W^\alpha \mathcal{O})$ at $k=3$ since $b_2+1=2$ for T$^{1,1}$.}
\label{tab:T11}
\end{table}

\subsection{\texorpdfstring{$\#6(\text{S}^2\times\text{S}^3)$}{\#6(S2xS3)}}

% del Pezzo info: https://www.fanography.info/delpezzos

As our final example, we consider $\#6(\text{S}^2\times\text{S}^3)$, which is a $\Uni1$ bundle over a $\dP6$ surface. In this case $I_B=1$. Using the various dualities, one finds the independent transverse Dolbeault cohomologies are\footnote{These were computed using Macaulay2~\cite{M2} by defining $\dP6$ as a smooth cubic surface in $\mathbb{P}^3$.}
\begin{equation}
\begin{split}
  \TVh00k & = \tfrac{1}{2} (3 s^2+3 s+2) \, [s\geq0] , \\
    \TVh10k & =(3 s^2-7)\,[s\geq2] + [s=2],\\
    \TVh11k & = (7-3|s|)\, [|s|\leq2] , \\
    \end{split}
\end{equation}
where $3k=s$ and $s$ takes integer values. Thus it is natural to write the Hilbert series as $\tilde{H}(t)=H(t^6)$, giving, for the undeformed case,
\begin{equation}
    H(t) = 1 + 4t + 10t^2+ 19t^3 + \ldots = \frac{1+t^3}{(1-t)^4} , 
\end{equation}
and the single-trace index is
\begin{equation}
    \Ist(t) = \frac{9t^6}{1-t^6} .
\end{equation}

Though we will not give the details here, the field theory can be described by a quiver \cite{Franco:2004rt}. There are $\TVh003=4$ exactly marginal superpotential deformations of the undeformed theory of the form $\tr\mathcal{O}_f$, which correspond to a choice of section $s$ of $K_B^{-1}$. Recall that the divisor defined by $s=0$ is a cubic in $\mathbb{P}^2$ fixed to pass through the six blown-up points. Since only the relative positions of the points are fixed, this indeed leaves four degrees of freedom in the choice of cubic. Infinitesimally, these deformations introduce three-form flux to the supergravity background. There are also $\TVH113=4$ exactly marginal deformations of the form $\tr\mathcal{O}_w$ corresponding, infinitesimally, to deformations of the Einstein metric on $\dP6$. Finally, there are $b_2+1=b_2(B)=7$ marginal deformations of the form $\tr W^\alpha W_\alpha \mathcal{O}$ that deform the gauge coupling constants. For the generic marginally deformed theory, the $f$ in $\eta=\dd f$ is the lift of the section of $K_B^{-1}$ on $\dP6$ to $\#6(\text{S}^2\times\text{S}^3)$, and counting the chiral operators gives, from \eqref{eq:etaH2-allk}, 
\begin{equation}
    \dim \etaH 2k = \TVindex k - [k\equiv_3 0]
        = 8 [k\equiv_3 0]  ,
\end{equation}
so that the Hilbert series \eqref{eq:Hilbert-def} is given by $\tilde{H}(t)=H(t^6)$ with 
\begin{equation}
    H(t) % &=\sum_k q^0_k t^k = \sum_k \bigl(\dim \etaH 2{k+3} - [k=0]\bigr) t^k \\
    = 1 + 8t + 8t^2 + 8t^3 + 8t^4 + \ldots
    =\frac{1+7t}{1-t} .
\end{equation}
In addition, we have again the general results $\etaH 1k=0$ and $\etaH 0k=[k\equiv_30]\bbC$. 

We could not find any direct calculation of the dimension of the cyclic homology of the non-commutative Calabi--Yau algebra $A$ for the deformation of $\dP6$, and so these can be regarded as a prediction for the form of $\rHC_n(A,k)$. The complete counting of scalar chiral perturbations due to deformations and vevs is given in Table \ref{tab:dP6}.

Note that for the general case of $\#n(\text{S}^2\times\text{S}^3)$, which are $\Uni1$ bundles over a $\dP n$ surface, the single-trace index is \cite{Eager:2012hx}
\begin{equation}
    \Ist(t) = \frac{(n+3)t^6}{1-t^6},
\end{equation}
giving the Hilbert series for the deformed theories as $\tilde{H}(t)=H(t^6)$ with
\begin{equation}
    H(t) % &=\sum_k q^0_k t^k = \sum_k \bigl(\dim \etaH 2{k+3} - [k=0]\bigr) t^k \\
    = 1 + (n+2)t + (n+2)t^2 + (n+2)t^3 + \ldots
    =\frac{1+(n+1)t}{1-t} .
\end{equation}

\begin{table}[t]
\centering
\renewcommand*{\arraystretch}{1.1}
\begin{tabular}{@{}l>{\raggedleft}p{1.2em}>{\raggedleft}p{1.2em}>{\raggedleft}p{1.2em}>{\raggedleft}p{1.2em}>{\raggedleft}p{1.2em}>{\raggedleft}p{1.2em}>{\raggedleft}p{1.2em}>{\raggedleft}p{1.2em}>{\raggedleft}p{1.2em}>{\raggedleft}p{1.2em}@{}}
\toprule
$k$ & $-12$ & $-9$ & $-6$ & $-3$ & $0$ & $3$ & $6$ & $9$ & $12$ & $15$ \tabularnewline
\midrule
$\tr(\mathcal{O})$ & &  &  &  & & $8$ & $8$ & $8$ & $8$ & $8$  \tabularnewline
$\tr(W^{2}\mathcal{O})$ &  &  &  &  &  & $7$ & $8$ & $8$ & $8$ & $8$  \tabularnewline
$\tr(\overline{\mathcal{O}})$ & $8$ & $8$ & $8$ & $8$ & $8$  \tabularnewline
$\tr(\overline{W^{2}\mathcal{O}})$ & $8$ & $8$ & $8$ & $8$ & $7$ \tabularnewline
\midrule
Total & $16$ & $16$ & $16$ & $16$ & $15$ & $15$ & $16$ & $16$ & $16$ & $16$ \tabularnewline
\bottomrule
\end{tabular}
\caption{Counting of operators in field theory dual to $\#6(\text{S}^2\times\text{S}^3)$ graded by $k$. Note that the number of $\tr(W_\alpha W^\alpha \mathcal{O})$ operators, except for the case of $k=0$, is the same as the number of $\tr(\mathcal{O})$ operators after a shift of $k$ by 3. There are seven modes for $\tr(W_\alpha W^\alpha \mathcal{O})$ at $k=3$ since $b_2+1=7$ for  $\#6(\text{S}^2\times\text{S}^3)$.}
\label{tab:dP6}
\end{table}

\subsection*{Acknowledgements}
ET was supported by an STFC PhD studentship while carrying out the work that led to this publication.

\bibliographystyle{utphys}
\bibliography{citations,extra}

\end{document}